\documentclass[
reprint,
superscriptaddress,
 amsmath,amssymb,
 aps,
]{revtex4-1}

\usepackage{graphicx}
\usepackage{xcolor}
\usepackage[normalem]{ulem}
\usepackage{braket}
\usepackage{tikz}
\usepackage{changes}
\usepackage{dcolumn}
\usepackage{bm}
\usepackage{siunitx}
\usepackage{microtype}
\usepackage{upgreek}
\usepackage{hyperref}
\hypersetup{
    colorlinks,
    linkcolor={blue!80!black},
    citecolor={blue!80!black},
    urlcolor={blue!80!black}
}

\definecolor{myblue}{RGB}{0, 112, 192}

\DeclareUnicodeCharacter{2212}{-}

\begin{document}

\author{L.~Banszerus}
\affiliation{Center for Quantum Devices, Niels Bohr Institute, University of Copenhagen, 2100 Copenhagen, Denmark}
\author{W. Marshall}
\affiliation{Center for Quantum Devices, Niels Bohr Institute, University of Copenhagen, 2100 Copenhagen, Denmark}
\affiliation{Department of Physics, University of Washington, Seattle, Washington 98195, USA}%
\author{C. W. Andersson}
\affiliation{Center for Quantum Devices, Niels Bohr Institute, University of Copenhagen, 2100 Copenhagen, Denmark}
\author{T. Lindemann}
\affiliation{Department of Physics and Astronomy, Purdue University, West Lafayette, Indiana 47907, USA}%
\affiliation{Birck Nanotechnology Center, Purdue University, West Lafayette, Indiana 47907, USA}
\author{M. J. Manfra}
\affiliation{Department of Physics and Astronomy, Purdue University, West Lafayette, Indiana 47907, USA}%
\affiliation{Birck Nanotechnology Center, Purdue University, West Lafayette, Indiana 47907, USA}
\affiliation{School of Electrical and Computer Engineering, Purdue University, West Lafayette, Indiana 47907, USA}
\affiliation{School of Materials Engineering, Purdue University, West Lafayette, Indiana 47907, USA}
\author{C. M. Marcus}
\affiliation{Center for Quantum Devices, Niels Bohr Institute, University of Copenhagen, 2100 Copenhagen, Denmark}
\affiliation{Department of Physics, University of Washington, Seattle, Washington 98195, USA}%
\affiliation{Materials Science and Engineering, University of Washington, Seattle, Washington 98195, USA}%
\author{S. Vaitiek\.{e}nas}
\affiliation{Center for Quantum Devices, Niels Bohr Institute, University of Copenhagen, 2100 Copenhagen, Denmark}%

\title{Voltage-controlled synthesis of higher harmonics in hybrid Josephson junction circuits}

\date{\today}

\keywords{Josephson junction, current-phase-relation, hybrid material}

\begin{abstract} 
We report measurements of the current-phase relation of two voltage-controlled semiconductor-superconductor hybrid Josephson junctions (JJs) in series. The two hybrid junctions behave similar to a single-mode JJ with effective transparency determined by the ratio of Josephson coupling strengths of the two junctions. Gate-voltage control of Josephson coupling (measured from switching currents) allows tuning of the harmonic content from sinusoidal, for asymmetric tuning, to highly nonsinusoidal, for symmetric tuning. The experimentally observed tunable harmonic content agrees with a model based on two conventional (sinusoidal) JJs in series.  
\end{abstract}

\maketitle
Josephson junctions (JJs) support a dissipationless supercurrent, $I(\varphi)$, that depends on the phase difference, $\varphi$, between the leads~\cite{Josephson1962Jul}.
In conventional superconducting tunnel junctions, sequential coherent transfer of Cooper pairs results in a sinusoidal current-phase relation (CPR)~\cite{Golubov2004Apr}.
In contrast, highly transparent JJs allow simultaneous coherent transport of $n$ Cooper pairs, yielding a nonsinusoidal CPR with higher harmonics, ${I(\varphi)=\sum\nolimits_{n=1}^\infty A_n \sin(n \varphi)}$~\cite{Ishii1970, Likharev1979, DellaRocca2007Sep, Murani2017Jul}.

Controlling the harmonic content of Josephson elements is crucial for advancing superconducting quantum technologies.
Circuits with tunable CPRs allow for simulating the emergence of quantum phases and their transitions~\cite{Kuzmanovski2023Dec, Maffi2024May}.
Higher harmonics can be used to minimize the charge dispersion of qubit states, thereby enhancing coherence times~\cite{Kringhoj2020Jun, Bargerbos2020Jun, Willsch2023Feb}, or to engineer double-well Josephson potentials realizing parity-protected qubits with suppressed relaxation and dephasing~\cite{Ioffe2002Dec, Doucot2012Jun, Bell2014Apr, deLange2015Sep, Smith2020Jan, Gyenis2021Sep, Schrade2022Jul}. 
Furthermore, nonsinusoidal Josephson circuits enable nonlinear superconducting elements~\cite{Hover2012Feb, Frattini2017May}, parametric amplifiers~\cite{Abdo2013Jul, Sivak2020Feb, Schrade2023Oct}, and nonreciprocal devices like Josephson diodes~\cite{Baumgartner2022Jan, Souto2022May, Ciaccia2023Apr, Bozkurt2023Jul}.

One of the main approaches for realizing nonsinusoidal Josephson elements involves circuits comprising multiple sinusoidal JJs.
In these circuits, the total phase difference is distributed across the individual JJs, leading to an overall anharmonic behavior~\cite{Golubov2004Apr, Barash2018Jun, Coraiola2023May, Matsuo2023Dec}.
The resulting CPRs depend on the Josephson energies, $E_\mathrm{J}$, of each junction but are largely independent of microscopic details.
One example is the Josephson rhombus with four JJs in a loop~\cite{Doucot2002May, Pop2008Sep, Gladchenko2009Jan}.

An alternative route to higher harmonics is based on individual highly-transparent JJs.
Such junctions are typically realized in superconductor-normal conductor-superconductor (SNS) structures with inherently nonsinusoidal CPRs, given by~\cite{Beenakker1991Dec}
\begin{equation}
    I(\varphi)=\sum_{i=1}^N\frac{e\Delta}{2\hbar}\frac{\tau_i\,\sin(\varphi)}{\sqrt{1-\tau_i\,\sin^2(\varphi/2)}}\,,
    \label{Beenakker}
\end{equation}
where $\Delta$ is the superconducting gap, $\tau_i$ is the transparency of the $i^{\rm th}$ mode, and $N$ is the number of modes.
Junctions with high transparency have been observed in various high-mobility platforms, including semiconducting nanowires~\cite{vanWoerkom2017Sep, Spanton2017Dec, Larsen2020Jul, Ueda2020Sep}, quantum wells~\cite{Nichele2020Jun, Ciaccia2023Jun, Valentini2023Jun, Leblanc2023Nov}, van-der-Waals materials~\cite{English2016Sep, Nanda2017, Portoles2022Nov}, and topological insulators~\cite{Sochnikov2015Feb, Gathak2018Jul}.
These systems allow for voltage-controlled JJs~\cite{vanDam2006Aug, Cleuziou2006Oct, Kjaergaard2017Mar}, whose CPRs depend on device-specific microscopic details affecting the transparencies of each contributing mode and thus the harmonic composition of the CPR. 

Here, we investigate hybrid Josephson elements composed of two semiconducting junctions in series with voltage-tunable harmonic content.
Previous theory showed that CPR of two sinusoidal JJs in series can be described by Eq.~\eqref{Beenakker}; see Ref.~\cite{Bozkurt2023Jul}.
In this case, $\Delta$ is replaced by the sum of the Josephson energies, $\sigma = E_\mathrm{J1}+E_\mathrm{J2}$, and the effective transparency is given by $\tau=4\rho/(1+\rho)^2$, where $\rho=E_\mathrm{J1}/E_\mathrm{J2}$.
We experimentally tune $\rho$ by changing the ratio of the two-junction switching currents, $I_\mathrm{SW1}/I_\mathrm{SW2}$, and demonstrate a controlled transition from sinusoidal to highly nonsinusoidal CPR, with substantial contributions from the first four harmonics.
Realizing the same harmonic content in a single JJ would require a mode with an intrinsic transparency $\tau_\mathrm{int} > 0.95$, which remains challenging to achieve in a controlled manner due to disorder-induced fluctuations in $E_\mathrm{J}$~\cite{Goffman2017Sep, Hart2019Aug}.

\begin{figure}[t]
    \includegraphics[width=\linewidth]{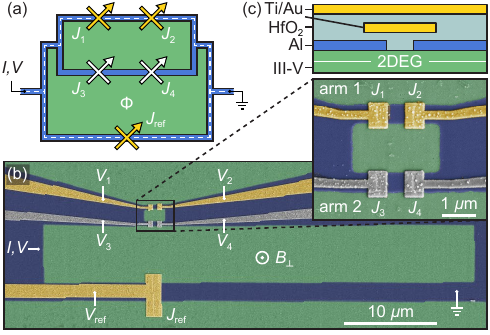}
    \caption{ 
    (a) Schematic of the measured device comprising two parallel arms, each with two voltage-tunable JJs in series, embedded in a loop with a reference junction for current-phase relation measurements.
    (b) Color-enhanced scanning electron micrograph of a reference device, taken before the deposition of the global top gate.
    The exposed semiconductor is green, and the epitaxial Al is blue.
    Colorized gate configuration highlights the setup for measuring the CPR of arm 1, with conducting (depleted) JJs under yellow (gray) gates.
    The inset shows a zoom-in on the two arms of the device.
    (c) Schematic cross-section of a junction showing the dual-gate configuration.
    The local gate (first layer) tunes the Josephson coupling, while the global gate (second layer, kept at $-1.3~V$ throughout the experiment) depletes the surrounding 2DEG.  
    }
    \label{f1}
\end{figure}

Devices were fabricated using an InAs two-dimensional electron gas (2DEG) heterostructure, proximitized by an epitaxial Al layer.
The Al film was lithographically patterned to form two parallel arms, each with two JJs in series, denoted $J_1$, $J_2$ in arm~1, and $J_3$, $J_4$ in arm~2.
Both arms were embedded in a loop with a reference junction, $J_\mathrm{ref}$, designed to have a much higher critical current compared to the other JJs; see Figs.~\ref{f1}(a) and \ref{f1}(b).
Individual junction Ti/Au top gates were metallized after atomic layer deposition of HfOx, allowing for independent electrostatic control.
A second set of HfOx dielectric and a global Ti/Au gate were fabricated to deplete the surrounding 2DEG.
The complete device stack is illustrated in Fig.~\ref{f1}(c).
The two arms showed similar results. We report representative data from both devices in the main text and present supporting data in the Supplemental Material~\cite{Supplement}.
Measurements were performed in a dilution refrigerator with a three-axis vector magnet and base temperature of 20~mK using standard ac lock-in techniques in a four-terminal configuration. 
Further details on wafer structure, sample fabrication, and measurements are given in the Supplemental Material~\cite{Supplement}.

\begin{figure}[t]
    \includegraphics[width=\linewidth]{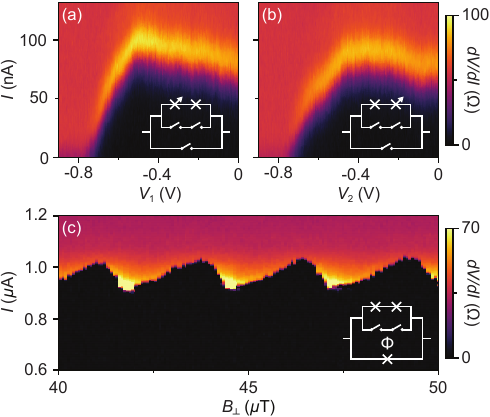}
     \caption{
    (a) Differential resistance, $dV/dI$, as a function of current bias, $I$, and $J_1$ gate voltage, $V_\mathrm{1}$, measured for arm~1 with open $J_2$ ($V_\mathrm{2} = 0$). Arm 2 and $J_\mathrm{ref}$ were pinched off ($V_\mathrm{4} = -1$~V, $V_{\rm ref} = - 1.5$~V).
    (b) Similar to (a) but taken as a function of $V_\mathrm{2}$, with $V_\mathrm{1}=0$.
    For both junctions, the switching current is maximal around $V_1,V_2 = -0.5$~V and is fully suppressed at $-0.8$~V. 
    (c) Resistance as a function of $I$ and flux-threading perpendicular magnetic field, $B_\perp$, measured with arm 1 and $J_{\rm ref}$ open, but arm 2 isolated.
    The skewed oscillations in the switching current suggest a highly nonsinusoidal current-phase relation.
    }
    \label{f2} 
\end{figure}

We begin by investigating the supercurrent transport through arm 1, while keeping arm 2 and $J_\mathrm{ref}$ pinched off.
Differential resistance, $dV/dI$, as a function of current bias, $I$, and gate voltage $V_1$ [Fig.~\ref{f2}(a)] or $V_2$ [Fig.~\ref{f2}(b)], shows switching current, $I_{\rm SW}$, that initially increases with decreasing gate voltage, reaching nearly 100~nA at $V_\mathrm{1}, V_\mathrm{2} \sim -0.5~V$, but then gets fully suppressed at around $-0.8~V$.
A comparison of the switching and retrapping currents shows no hysteresis, indicating overdamped junctions~\cite{Supplement}.
To measure the CPR of arm 1, we open $J_{\rm ref}$ by setting $V_\mathrm{ref} = 0$ and apply a flux-threading perpendicular magnetic field, $B_\perp$.
The measured $dV/dI$ reveals a periodic switching current of $\sim1~\mu$A, modulated by nonsinusoidal oscillations with an amplitude of $\sim 100$~nA and a period of $\sim 2.5~\mu$T [see Fig.~\ref{f2}(c)].
We attribute the large background $I_{\rm SW}$ to $J_\mathrm{ref}$ and the oscillatory component, $\Delta I_{\rm SW}$, to the flux-induced modulation of phase difference across the two JJs in arm~1.

\begin{figure}[t]
    \includegraphics[width=\linewidth]{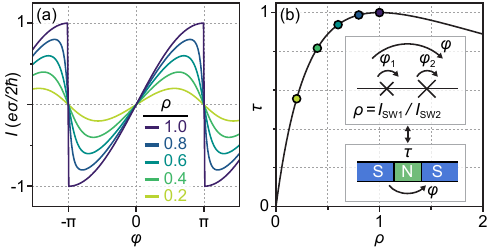}
    \caption{
    (a) Theoretical current-phase relation of two sinusoidal Josephson junctions in series [Eq.~\eqref{2JJ}] for various ratios of the Josephson energies $\rho=E_{\rm J1}/E_{\rm J2}$.
    The relation evolves from sinusoidal to highly nonsinusoidal as $\rho$ approaches 1.
    (b)~Mapping between $\rho$ and the transparency of a single-mode junction, $\tau$. 
     }
    \label{f3}
\end{figure}

To better understand the origin of the nonsinusoidal CPR, we model the measured Josephson elements as two sinusoidal junctions in series, following Ref.~\cite{Bozkurt2023Jul}.
The total energy of the system is $E_\mathrm{tot}=E_\mathrm{J1}[1-\cos(\varphi_1)]+E_\mathrm{J2}[1-\cos(\varphi_2)]$, with the total phase differences across both JJs defined by the external phase bias $\varphi=\varphi_1+\varphi_2$~\cite{Golubov2004Apr, Barash2018Jun}.
Minimizing the total energy under this boundary condition yields the phase-dependent ground-state energy $E_\mathrm{0}(\varphi)=-\sigma\sqrt{1-\frac{4\rho}{(1+\rho)^2}\sin^2(\varphi/2)}$.
The corresponding CPR is given by
\begin{equation}
\begin{split}
 I(\varphi) & =\frac{2e}{\hbar}\frac{\partial E_\mathrm{0}}{\partial \varphi}\\
 & =\frac{e\sigma}{2\hbar}\frac{4\rho}{(1+\rho)^2}\frac{\sin(\varphi)}{\sqrt{1-\frac{4\rho}{(1+\rho)^2}\sin^2(\varphi/2)}}
 \label{2JJ}
\end{split}
\end{equation}
We note that Eq.~(\ref{2JJ}) has the same functional form as the expression for CPR of a transparent single-mode junction described by Eq.~\eqref{Beenakker}, where $\sigma$ acts as $\Delta$ and the term $4\rho/(1+\rho)^2$ serves as $\tau$.
The symmetrized case ($\rho=1$) corresponds to one fully transmissive mode ($\tau = 1$), with highly nonsinusoidal CPR [Fig.~\ref{f3}(a)].
Away from the symmetry point, the effective transparency decreases but remains $\gtrsim 0.9$ for $\rho$ between 0.5 to 2 [Fig.~\ref{f3}(b)].
In the limit of $\rho \gg 1$ and $\rho \ll 1$, the CPR of the double JJ approaches the CPR of the single JJ with the smaller $E_\mathrm{J}$.
We find that a purely sinusoidal CPR of the individual JJs is a good approximation as long as $4\rho/(1+\rho)^2$ is larger than the intrinsic transparencies of the JJs. 
The case of nonsinusoidal JJs is studied numerically in the Supplemental Material~\cite{Supplement}.

\begin{figure}[t]
    \includegraphics[width=\linewidth]{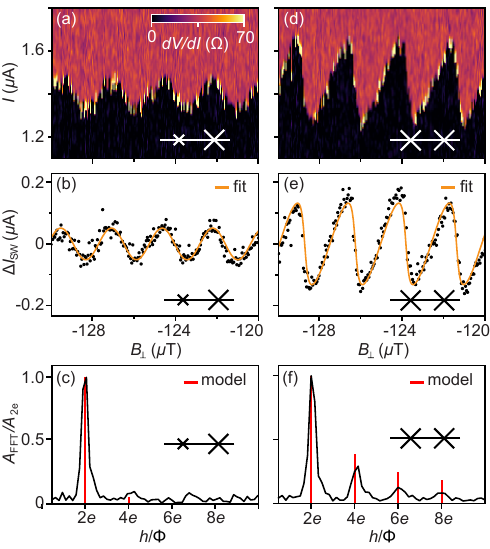}
    \caption{
    (a) Differential resistance, $dV/dI$, as a function of current bias, $I$, and perpendicular magnetic field, $B_\perp$, for arm~2 in the desymmetrized regime ($V_3=0$, $V_4=-0.76$~V), with open $J_{\rm ref}$ ($V_\mathrm{ref}=0$) but pinched-off arm~1 ($V_1=-1$~V, $V_2=-1$~V).
    (b)~Switching-current oscillations, $\Delta I_\mathrm{SW}$, inferred from data in (a).
    Fit to Eq.~(\ref{2JJ}) yields $\rho=0.02$, corresponding to an effective $\tau=0.09$.
    (c)~Fast Fourier transform amplitude of $\Delta I_\mathrm{SW}$ from (b), in the range of $B_\perp=-140$ to $-110~\mu$T.
    The red bars indicate the expected amplitude of the harmonics calculated from the ratio of the switching currents of $J_3$ and $J_4$.    
    \mbox{(d)--(f)}~Same as (a)--(c) but for the symmetrized case with $V_3=0$ and $V_4=-0.56$~V.
    The fit to Eq.~(\ref{2JJ}) in (e), yields $\rho=0.63$, corresponding to an effective $\tau=0.95$. 
    }
    \label{f4}
\end{figure}

To test the model experimentally, we investigate the CPR of arm 2 for different junction configurations
(analogous data from arm 1 is presented in the Supplemental Material~\cite{Supplement}). 
In the desymmetrized case, with highly conductive $J_3$ and nearly depleted $J_4$ ($V_3=0$, $V_4=-0.76$~V), the switching current exhibits roughly sinusoidal modulations,~see Fig.~\ref{f4}(a).
We attribute the additional fine modulations in the switching current to the switching statistics of the superconducting circuit.
For a quantitative analysis, we extract the oscillatory component of the switching current, $\Delta I_\mathrm{SW}$, corresponding to the CPR of arm 2  [Fig.~\ref{f4}(b)].
Fitting these data to Eq.~\eqref{2JJ} yields $\rho=0.02\pm0.02$ [orange curve in Fig.~\ref{f4}(b)].
Taking a fast Fourier transform of the measured $\Delta I_\mathrm{SW}$ reveals the harmonic content of the CPR, characterized by a prominent peak at the first harmonic ($2e$) and nearly absent higher-order components [Fig.~\ref{f4}(c)].
The observed Fourier spectrum agrees well with the expected harmonic amplitudes [red bars in Fig.~\ref{f4}(c)], determined by the independently measured ratio of the switching currents, $\rho = I_\mathrm{SW3}/I_\mathrm{SW4}=0.08\pm0.07$, with the confidence interval based on the width of $I_{\rm SW}$ features.

In the symmetrized scenario, with both $J_3$ and $J_4$ open ($V_3=0$, $V_4=-0.56~V$), the oscillatory component of $I_\mathrm{SW}$ displays a highly nonsinusoidal, sawtooth-like behavior, see Fig.~\ref{f4}(d).
A fit of Eq.~\eqref{2JJ} to the extracted $\Delta I_\mathrm{SW}$ yields $\rho=0.63\pm0.01$ [Fig.~\ref{f4}(e)].
Mapping this value to a single-mode JJ corresponds to a transparency of $\tau=0.95\pm0.05$.
The Fourier transform of $\Delta I_\mathrm{SW}$ reveals considerable contributions from the first four harmonics ($2e$, $4e$, $6e$, and $8e$), in a good agreement with the expected harmonic amplitudes based on independently measured, $\rho=I_\mathrm{SW3}/I_\mathrm{SW4}=0.82\pm0.02$; see Fig.~\ref{f4}(f).
Arm 1 shows similar behavior for symmetrized and desymmetrized configurations~\cite{Supplement}.

\begin{figure}[t]
    \includegraphics[width=\linewidth]{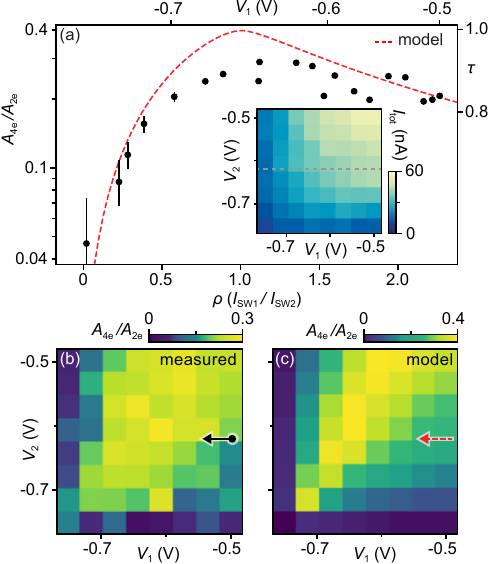}
    \caption{
    (a) Ratio of second to first harmonic amplitudes, $A_\mathrm{4e}/A_\mathrm{2e}$, determined from fits to CPR, plotted against independently measured $\rho = I_{\rm SW1}/I_{\rm SW2}$ for arm 1.
    Error bars indicate uncertainties in the fits to Eq.~\eqref{2JJ}.
    Data were taken with varying $V_1$ between $-0.75$ and $-0.5$~V and fixed $V_2=-0.62$~V.
    The dashed red curve is the expected $A_\mathrm{4e}/A_\mathrm{2e}$ based on the measured switching currents.
    Inset: gate-voltage evolution of the total switching current, $I_{\rm tot}$, of arm 1 determined from fits to CPRs.   
    (b) Measured and (c) expected $A_\mathrm{4e}/A_\mathrm{2e}$ ratio as a function of $V_1$ and $V_2$.
    }
    \label{f5}
\end{figure}
 
Finally, we demonstrate experimentally a deterministic control of harmonic content by investigating the CPR of arm 1 as a function of $V_1$, keeping $V_2$ fixed.
To quantify the nonsinusoidal nature of the CPR, we Fourier transform the measured $\Delta I_{\rm SW}$ and study the ratio between the second and first harmonic, $A_\mathrm{4e}/A_\mathrm{2e}$.
Figure~\ref{f5}(a) displays a parametric plot of $A_\mathrm{4e}/A_\mathrm{2e}$ against the independently measured $\rho = I_{\rm SW1}/I_{\rm SW2}$, with $V_1$ varying from $-0.75$ to $-0.5$~V and $V_2$ set to $-0.62$~V.
The measured  $A_\mathrm{4e}/A_\mathrm{2e}$ follows the expected behavior based on Eq.~\eqref{2JJ} [dashed red curve in Fig.~\ref{f5}(a)], reaching a value of $A_\mathrm{4e}/A_\mathrm{2e} \sim 0.3$ around $\rho=1$, lower than the theoretical maximum of 0.4.
Arm 2 shows a qualitatively similar behavior, with a weaker $\rho$-dependence of $A_\mathrm{4e}/A_\mathrm{2e}$, which we attribute to a non-negligible intrinsic junction transparency (see Supplemental Material~\cite{Supplement}).

It is unlikely that the observed discrepancy between measured and expected $A_\mathrm{4e}/A_\mathrm{2e}$ near $\rho = 1$ is due to thermal smearing, as $k_\mathrm{B}T\sim 2~\mu$eV is about two orders of magnitude smaller than the Josephson energy $\hbar\,\Delta I_\mathrm{SW}/2e\sim 0.2~$meV.
To examine the self-biasing as a possible explanation, we estimate the kinetic inductance of the superconducting loop to be $L_\mathrm{K}\sim 300$~pH~\cite{Annunziata2010Oct}.
This indicates that the nonlinearities between the total and external fluxes, $\Phi=\Phi_\mathrm{ext}-L_\mathrm{K} \Delta I_{\rm SW}(\varphi)$, can amount to $\sim 1\%$ of the superconducting flux quantum, $\Phi_0$~\cite{Nichele2020Jun}.
The estimated geometric inductance of the loop, $L_\mathrm{G}\sim 40$~pH, is considerably smaller and thus negligible compared to $L_\mathrm{K}$.
Instead, we suspect that the measured values of $A_\mathrm{4e}/A_\mathrm{2e}$ are limited by flux noise originating from the relatively large loop area and wide leads, resulting in a smearing of the CPR.\\

A map of $A_\mathrm{4e}/A_\mathrm{2e}$, measured as a function of $V_1$ and $V_2$, reveals that the ratio of the first two harmonics is maximal along the diagonal where $V_1 \sim V_2$, with the stability range expanding at less negative voltage, see Fig.~\ref{f5}(b).
The data is in good qualitative agreement with the expected $A_\mathrm{4e}/A_\mathrm{2e}$ based on Eq.~\eqref{2JJ} and the independently measured $I_\mathrm{SW1}(V_1)$ and $I_\mathrm{SW2}(V_2)$ [Fig.~\ref{f5}(c)].
We note that away from pinch-off, $A_\mathrm{4e}/A_\mathrm{2e}$ remains approximately constant for symmetrized arm 1, whereas its total switching current, $I_\mathrm{tot}=\sigma \rho/\Phi_0 (1+\rho)^2$, continues increasing with the gate voltages; see the inset in Fig.~\ref{f5}(a).
This highlights the unique feature of the investigated hybrid Josephson elements, namely the independent voltage-control of the harmonics content~(characterized by $A_\mathrm{4e}/A_\mathrm{2e}$) and the effective Josephson energy~(represented by $I_\mathrm{tot}$).\\

In summary, we have demonstrated an experimental method to deterministically synthesize tunable and highly nonsinusoidal Josephson elements consisting of two junctions in series, implemented in an InAs/Al hybrid heterostructure.
The approach combines the advantages of voltage control and resilience against microscopic device-to-device variations.
By controlling the switching current of each junction, the effective Josephson energy and harmonic content of the Josephson element can be tuned independently.
Our findings pave a new way for the realization of alternative superconducting qubits with improved coherence times, such as charge-insensitive qubits based on high effective transmission~\cite{Kringhoj2020Jun, Bargerbos2020Jun} or parity-protected qubits in a $\cos(2\varphi)$ potential~\cite{Smith2020Jan, Schrade2022Jul}.
The \textit{in-situ} tunability of the Josephson potential enables advanced qubits operation schemes~\cite{Krojer2024Apr}, tunable parametric amplifiers~\cite{Abdo2013Jul, Sivak2020Feb, Schrade2023Oct}, nonlinear Josephson circuits~\cite{Pillet2019Aug, Melo2022Jan, Matute2023Dec}, and nonreciprocal Josephson diodes~\cite{Souto2022May, Bozkurt2023Jul, Zhang2022Nov}.
\\

We thank A.~C.~C.~Drachmann and S. Sasmal for assistance with the device fabrication and M. Kjaergaard for discussions. We acknowledge support from research grants (Projects No. 43951, No. 50334, and No. 53097) 
from VILLUM FONDEN, the Danish National Research Foundation, and the European Research Council (Grant Agreement No. 856526).

\bibliography{Literature}

\begin{thebibliography}{59}%
\makeatletter
\providecommand \@ifxundefined [1]{%
 \@ifx{#1\undefined}
}%
\providecommand \@ifnum [1]{%
 \ifnum #1\expandafter \@firstoftwo
 \else \expandafter \@secondoftwo
 \fi
}%
\providecommand \@ifx [1]{%
 \ifx #1\expandafter \@firstoftwo
 \else \expandafter \@secondoftwo
 \fi
}%
\providecommand \natexlab [1]{#1}%
\providecommand \enquote  [1]{``#1''}%
\providecommand \bibnamefont  [1]{#1}%
\providecommand \bibfnamefont [1]{#1}%
\providecommand \citenamefont [1]{#1}%
\providecommand \href@noop [0]{\@secondoftwo}%
\providecommand \href [0]{\begingroup \@sanitize@url \@href}%
\providecommand \@href[1]{\@@startlink{#1}\@@href}%
\providecommand \@@href[1]{\endgroup#1\@@endlink}%
\providecommand \@sanitize@url [0]{\catcode `\\12\catcode `\$12\catcode `\&12\catcode `\#12\catcode `\^12\catcode `\_12\catcode `\%12\relax}%
\providecommand \@@startlink[1]{}%
\providecommand \@@endlink[0]{}%
\providecommand \url  [0]{\begingroup\@sanitize@url \@url }%
\providecommand \@url [1]{\endgroup\@href {#1}{\urlprefix }}%
\providecommand \urlprefix  [0]{URL }%
\providecommand \Eprint [0]{\href }%
\providecommand \doibase [0]{http://dx.doi.org/}%
\providecommand \selectlanguage [0]{\@gobble}%
\providecommand \bibinfo  [0]{\@secondoftwo}%
\providecommand \bibfield  [0]{\@secondoftwo}%
\providecommand \translation [1]{[#1]}%
\providecommand \BibitemOpen [0]{}%
\providecommand \bibitemStop [0]{}%
\providecommand \bibitemNoStop [0]{.\EOS\space}%
\providecommand \EOS [0]{\spacefactor3000\relax}%
\providecommand \BibitemShut  [1]{\csname bibitem#1\endcsname}%
\let\auto@bib@innerbib\@empty
\bibitem [{\citenamefont {Josephson}(1962)}]{Josephson1962Jul}%
  \BibitemOpen
  \bibfield  {author} {\bibinfo {author} {\bibfnamefont {B.~D.}\ \bibnamefont {Josephson}},\ }\href {\doibase 10.1016/0031-9163(62)91369-0} {\bibfield  {journal} {\bibinfo  {journal} {Phys. Lett.}\ }\textbf {\bibinfo {volume} {1}},\ \bibinfo {pages} {251} (\bibinfo {year} {1962})}\BibitemShut {NoStop}%
\bibitem [{\citenamefont {Golubov}\ \emph {et~al.}(2004)\citenamefont {Golubov}, \citenamefont {Kupriyanov},\ and\ \citenamefont {Il{'}ichev}}]{Golubov2004Apr}%
  \BibitemOpen
  \bibfield  {author} {\bibinfo {author} {\bibfnamefont {A.~A.}\ \bibnamefont {Golubov}}, \bibinfo {author} {\bibfnamefont {M.~{\relax Yu}.}\ \bibnamefont {Kupriyanov}}, \ and\ \bibinfo {author} {\bibfnamefont {E.}~\bibnamefont {Il{'}ichev}},\ }\href {\doibase 10.1103/RevModPhys.76.411} {\bibfield  {journal} {\bibinfo  {journal} {Rev. Mod. Phys.}\ }\textbf {\bibinfo {volume} {76}},\ \bibinfo {pages} {411} (\bibinfo {year} {2004})}\BibitemShut {NoStop}%
\bibitem [{\citenamefont {Ishii}(1970)}]{Ishii1970}%
  \BibitemOpen
  \bibfield  {author} {\bibinfo {author} {\bibfnamefont {C.}~\bibnamefont {Ishii}},\ }\href {\doibase 10.1143/ptp.44.1525} {\bibfield  {journal} {\bibinfo  {journal} {Prog. Theor. Phys.}\ }\textbf {\bibinfo {volume} {44}},\ \bibinfo {pages} {1525} (\bibinfo {year} {1970})}\BibitemShut {NoStop}%
\bibitem [{\citenamefont {Likharev}(1979)}]{Likharev1979}%
  \BibitemOpen
  \bibfield  {author} {\bibinfo {author} {\bibfnamefont {K.~K.}\ \bibnamefont {Likharev}},\ }\href {\doibase 10.1103/revmodphys.51.101} {\bibfield  {journal} {\bibinfo  {journal} {Rev. Mod. Phys.}\ }\textbf {\bibinfo {volume} {51}},\ \bibinfo {pages} {101} (\bibinfo {year} {1979})}\BibitemShut {NoStop}%
\bibitem [{\citenamefont {Della~Rocca}\ \emph {et~al.}(2007)\citenamefont {Della~Rocca}, \citenamefont {Chauvin}, \citenamefont {Huard}, \citenamefont {Pothier}, \citenamefont {Esteve},\ and\ \citenamefont {Urbina}}]{DellaRocca2007Sep}%
  \BibitemOpen
  \bibfield  {author} {\bibinfo {author} {\bibfnamefont {M.~L.}\ \bibnamefont {Della~Rocca}}, \bibinfo {author} {\bibfnamefont {M.}~\bibnamefont {Chauvin}}, \bibinfo {author} {\bibfnamefont {B.}~\bibnamefont {Huard}}, \bibinfo {author} {\bibfnamefont {H.}~\bibnamefont {Pothier}}, \bibinfo {author} {\bibfnamefont {D.}~\bibnamefont {Esteve}}, \ and\ \bibinfo {author} {\bibfnamefont {C.}~\bibnamefont {Urbina}},\ }\href {\doibase 10.1103/PhysRevLett.99.127005} {\bibfield  {journal} {\bibinfo  {journal} {Phys. Rev. Lett.}\ }\textbf {\bibinfo {volume} {99}},\ \bibinfo {pages} {127005} (\bibinfo {year} {2007})}\BibitemShut {NoStop}%
\bibitem [{\citenamefont {Murani}\ \emph {et~al.}(2017)\citenamefont {Murani}, \citenamefont {Kasumov}, \citenamefont {Sengupta}, \citenamefont {Kasumov}, \citenamefont {Volkov}, \citenamefont {Khodos}, \citenamefont {Brisset}, \citenamefont {Delagrange}, \citenamefont {Chepelianskii}, \citenamefont {Deblock}, \citenamefont {Bouchiat},\ and\ \citenamefont {Gu{\ifmmode\acute{e}\else\'{e}\fi}ron}}]{Murani2017Jul}%
  \BibitemOpen
  \bibfield  {author} {\bibinfo {author} {\bibfnamefont {A.}~\bibnamefont {Murani}}, \bibinfo {author} {\bibfnamefont {A.}~\bibnamefont {Kasumov}}, \bibinfo {author} {\bibfnamefont {S.}~\bibnamefont {Sengupta}}, \bibinfo {author} {\bibfnamefont {Y.~A.}\ \bibnamefont {Kasumov}}, \bibinfo {author} {\bibfnamefont {V.~T.}\ \bibnamefont {Volkov}}, \bibinfo {author} {\bibfnamefont {I.~I.}\ \bibnamefont {Khodos}}, \bibinfo {author} {\bibfnamefont {F.}~\bibnamefont {Brisset}}, \bibinfo {author} {\bibfnamefont {R.}~\bibnamefont {Delagrange}}, \bibinfo {author} {\bibfnamefont {A.}~\bibnamefont {Chepelianskii}}, \bibinfo {author} {\bibfnamefont {R.}~\bibnamefont {Deblock}}, \bibinfo {author} {\bibfnamefont {H.}~\bibnamefont {Bouchiat}}, \ and\ \bibinfo {author} {\bibfnamefont {S.}~\bibnamefont {Gu{\ifmmode\acute{e}\else\'{e}\fi}ron}},\ }\href {\doibase 10.1038/ncomms15941} {\bibfield  {journal} {\bibinfo  {journal} {Nat. Commun.}\ }\textbf {\bibinfo {volume} {8}},\ \bibinfo {pages} {1} (\bibinfo {year}
  {2017})}\BibitemShut {NoStop}%
\bibitem [{\citenamefont {Kuzmanovski}\ \emph {et~al.}()\citenamefont {Kuzmanovski}, \citenamefont {Souto}, \citenamefont {Wong},\ and\ \citenamefont {Balatsky}}]{Kuzmanovski2023Dec}%
  \BibitemOpen
  \bibfield  {author} {\bibinfo {author} {\bibfnamefont {D.}~\bibnamefont {Kuzmanovski}}, \bibinfo {author} {\bibfnamefont {R.~S.}\ \bibnamefont {Souto}}, \bibinfo {author} {\bibfnamefont {P.~J.}\ \bibnamefont {Wong}}, \ and\ \bibinfo {author} {\bibfnamefont {A.~V.}\ \bibnamefont {Balatsky}},\ }\href@noop {} {\bibinfo  {journal} {\href{https://arxiv.org/abs/2312.03456}{arXiv:2312.03456}}\ }\BibitemShut {NoStop}%
\bibitem [{\citenamefont {Maffi}\ \emph {et~al.}(2024)\citenamefont {Maffi}, \citenamefont {Tausendpfund}, \citenamefont {Rizzi},\ and\ \citenamefont {Burrello}}]{Maffi2024May}%
  \BibitemOpen
\bibfield  {journal} {  }\bibfield  {author} {\bibinfo {author} {\bibfnamefont {L.}~\bibnamefont {Maffi}}, \bibinfo {author} {\bibfnamefont {N.}~\bibnamefont {Tausendpfund}}, \bibinfo {author} {\bibfnamefont {M.}~\bibnamefont {Rizzi}}, \ and\ \bibinfo {author} {\bibfnamefont {M.}~\bibnamefont {Burrello}},\ }\href {\doibase 10.1103/PhysRevLett.132.226502} {\bibfield  {journal} {\bibinfo  {journal} {Phys. Rev. Lett.}\ }\textbf {\bibinfo {volume} {132}},\ \bibinfo {pages} {226502} (\bibinfo {year} {2024})}\BibitemShut {NoStop}%
\bibitem [{\citenamefont {Kringh{\o}j}\ \emph {et~al.}(2020)\citenamefont {Kringh{\o}j}, \citenamefont {van Heck}, \citenamefont {Larsen}, \citenamefont {Erlandsson}, \citenamefont {Sabonis}, \citenamefont {Krogstrup}, \citenamefont {Casparis}, \citenamefont {Petersson},\ and\ \citenamefont {Marcus}}]{Kringhoj2020Jun}%
  \BibitemOpen
  \bibfield  {author} {\bibinfo {author} {\bibfnamefont {A.}~\bibnamefont {Kringh{\o}j}}, \bibinfo {author} {\bibfnamefont {B.}~\bibnamefont {van Heck}}, \bibinfo {author} {\bibfnamefont {T.~W.}\ \bibnamefont {Larsen}}, \bibinfo {author} {\bibfnamefont {O.}~\bibnamefont {Erlandsson}}, \bibinfo {author} {\bibfnamefont {D.}~\bibnamefont {Sabonis}}, \bibinfo {author} {\bibfnamefont {P.}~\bibnamefont {Krogstrup}}, \bibinfo {author} {\bibfnamefont {L.}~\bibnamefont {Casparis}}, \bibinfo {author} {\bibfnamefont {K.~D.}\ \bibnamefont {Petersson}}, \ and\ \bibinfo {author} {\bibfnamefont {C.~M.}\ \bibnamefont {Marcus}},\ }\href {\doibase 10.1103/PhysRevLett.124.246803} {\bibfield  {journal} {\bibinfo  {journal} {Phys. Rev. Lett.}\ }\textbf {\bibinfo {volume} {124}},\ \bibinfo {pages} {246803} (\bibinfo {year} {2020})}\BibitemShut {NoStop}%
\bibitem [{\citenamefont {Bargerbos}\ \emph {et~al.}(2020)\citenamefont {Bargerbos}, \citenamefont {Uilhoorn}, \citenamefont {Yang}, \citenamefont {Krogstrup}, \citenamefont {Kouwenhoven}, \citenamefont {de~Lange}, \citenamefont {van Heck},\ and\ \citenamefont {Kou}}]{Bargerbos2020Jun}%
  \BibitemOpen
  \bibfield  {author} {\bibinfo {author} {\bibfnamefont {A.}~\bibnamefont {Bargerbos}}, \bibinfo {author} {\bibfnamefont {W.}~\bibnamefont {Uilhoorn}}, \bibinfo {author} {\bibfnamefont {C.-K.}\ \bibnamefont {Yang}}, \bibinfo {author} {\bibfnamefont {P.}~\bibnamefont {Krogstrup}}, \bibinfo {author} {\bibfnamefont {L.~P.}\ \bibnamefont {Kouwenhoven}}, \bibinfo {author} {\bibfnamefont {G.}~\bibnamefont {de~Lange}}, \bibinfo {author} {\bibfnamefont {B.}~\bibnamefont {van Heck}}, \ and\ \bibinfo {author} {\bibfnamefont {A.}~\bibnamefont {Kou}},\ }\href {\doibase 10.1103/PhysRevLett.124.246802} {\bibfield  {journal} {\bibinfo  {journal} {Phys. Rev. Lett.}\ }\textbf {\bibinfo {volume} {124}},\ \bibinfo {pages} {246802} (\bibinfo {year} {2020})}\BibitemShut {NoStop}%
\bibitem [{\citenamefont {Willsch}\ \emph {et~al.}(2024)\citenamefont {Willsch}, \citenamefont {Rieger}, \citenamefont {Winkel}, \citenamefont {Willsch}, \citenamefont {Dickel}, \citenamefont {Krause}, \citenamefont {Ando}, \citenamefont {Lescanne}, \citenamefont {Leghtas}, \citenamefont {Bronn}, \citenamefont {Deb}, \citenamefont {Lanes}, \citenamefont {Minev}, \citenamefont {Dennig}, \citenamefont {Geisert}, \citenamefont {G{\ifmmode\ddot{u}\else\"{u}\fi}nzler}, \citenamefont {Ihssen}, \citenamefont {Paluch}, \citenamefont {Reisinger}, \citenamefont {Hanna}, \citenamefont {Bae}, \citenamefont {Sch{\ifmmode\ddot{u}\else\"{u}\fi}ffelgen}, \citenamefont {Gr{\ifmmode\ddot{u}\else\"{u}\fi}tzmacher}, \citenamefont {Buimaga-Iarinca}, \citenamefont {Morari}, \citenamefont {Wernsdorfer}, \citenamefont {DiVincenzo}, \citenamefont {Michielsen}, \citenamefont {Catelani},\ and\ \citenamefont {Pop}}]{Willsch2023Feb}%
  \BibitemOpen
  \bibfield  {author} {\bibinfo {author} {\bibfnamefont {D.}~\bibnamefont {Willsch}}, \bibinfo {author} {\bibfnamefont {D.}~\bibnamefont {Rieger}}, \bibinfo {author} {\bibfnamefont {P.}~\bibnamefont {Winkel}}, \bibinfo {author} {\bibfnamefont {M.}~\bibnamefont {Willsch}}, \bibinfo {author} {\bibfnamefont {C.}~\bibnamefont {Dickel}}, \bibinfo {author} {\bibfnamefont {J.}~\bibnamefont {Krause}}, \bibinfo {author} {\bibfnamefont {Y.}~\bibnamefont {Ando}}, \bibinfo {author} {\bibfnamefont {R.}~\bibnamefont {Lescanne}}, \bibinfo {author} {\bibfnamefont {Z.}~\bibnamefont {Leghtas}}, \bibinfo {author} {\bibfnamefont {N.~T.}\ \bibnamefont {Bronn}}, \bibinfo {author} {\bibfnamefont {P.}~\bibnamefont {Deb}}, \bibinfo {author} {\bibfnamefont {O.}~\bibnamefont {Lanes}}, \bibinfo {author} {\bibfnamefont {Z.~K.}\ \bibnamefont {Minev}}, \bibinfo {author} {\bibfnamefont {B.}~\bibnamefont {Dennig}}, \bibinfo {author} {\bibfnamefont {S.}~\bibnamefont {Geisert}}, \bibinfo {author} {\bibfnamefont {S.}~\bibnamefont
  {G{\ifmmode\ddot{u}\else\"{u}\fi}nzler}}, \bibinfo {author} {\bibfnamefont {S.}~\bibnamefont {Ihssen}}, \bibinfo {author} {\bibfnamefont {P.}~\bibnamefont {Paluch}}, \bibinfo {author} {\bibfnamefont {T.}~\bibnamefont {Reisinger}}, \bibinfo {author} {\bibfnamefont {R.}~\bibnamefont {Hanna}}, \bibinfo {author} {\bibfnamefont {J.~H.}\ \bibnamefont {Bae}}, \bibinfo {author} {\bibfnamefont {P.}~\bibnamefont {Sch{\ifmmode\ddot{u}\else\"{u}\fi}ffelgen}}, \bibinfo {author} {\bibfnamefont {D.}~\bibnamefont {Gr{\ifmmode\ddot{u}\else\"{u}\fi}tzmacher}}, \bibinfo {author} {\bibfnamefont {L.}~\bibnamefont {Buimaga-Iarinca}}, \bibinfo {author} {\bibfnamefont {C.}~\bibnamefont {Morari}}, \bibinfo {author} {\bibfnamefont {W.}~\bibnamefont {Wernsdorfer}}, \bibinfo {author} {\bibfnamefont {D.~P.}\ \bibnamefont {DiVincenzo}}, \bibinfo {author} {\bibfnamefont {K.}~\bibnamefont {Michielsen}}, \bibinfo {author} {\bibfnamefont {G.}~\bibnamefont {Catelani}}, \ and\ \bibinfo {author} {\bibfnamefont {I.~M.}\ \bibnamefont {Pop}},\
  }\href {\doibase 10.1038/s41567-024-02400-8} {\bibfield  {journal} {\bibinfo  {journal} {Nat. Phys.}\ ,\ \bibinfo {pages} {1}} (\bibinfo {year} {2024})}\BibitemShut {NoStop}%
\bibitem [{\citenamefont {Ioffe}\ and\ \citenamefont {Feigel{'}man}(2002)}]{Ioffe2002Dec}%
  \BibitemOpen
  \bibfield  {author} {\bibinfo {author} {\bibfnamefont {L.~B.}\ \bibnamefont {Ioffe}}\ and\ \bibinfo {author} {\bibfnamefont {M.~V.}\ \bibnamefont {Feigel{'}man}},\ }\href {\doibase 10.1103/PhysRevB.66.224503} {\bibfield  {journal} {\bibinfo  {journal} {Phys. Rev. B}\ }\textbf {\bibinfo {volume} {66}},\ \bibinfo {pages} {224503} (\bibinfo {year} {2002})}\BibitemShut {NoStop}%
\bibitem [{\citenamefont {Dou{\ifmmode\mbox{\c{c}}\else\c{c}\fi}ot}\ and\ \citenamefont {Ioffe}(2012)}]{Doucot2012Jun}%
  \BibitemOpen
  \bibfield  {author} {\bibinfo {author} {\bibfnamefont {B.}~\bibnamefont {Dou{\ifmmode\mbox{\c{c}}\else\c{c}\fi}ot}}\ and\ \bibinfo {author} {\bibfnamefont {L.~B.}\ \bibnamefont {Ioffe}},\ }\href {\doibase 10.1088/0034-4885/75/7/072001} {\bibfield  {journal} {\bibinfo  {journal} {Rep. Prog. Phys.}\ }\textbf {\bibinfo {volume} {75}},\ \bibinfo {pages} {072001} (\bibinfo {year} {2012})}\BibitemShut {NoStop}%
\bibitem [{\citenamefont {Bell}\ \emph {et~al.}(2014)\citenamefont {Bell}, \citenamefont {Paramanandam}, \citenamefont {Ioffe},\ and\ \citenamefont {Gershenson}}]{Bell2014Apr}%
  \BibitemOpen
  \bibfield  {author} {\bibinfo {author} {\bibfnamefont {M.~T.}\ \bibnamefont {Bell}}, \bibinfo {author} {\bibfnamefont {J.}~\bibnamefont {Paramanandam}}, \bibinfo {author} {\bibfnamefont {L.~B.}\ \bibnamefont {Ioffe}}, \ and\ \bibinfo {author} {\bibfnamefont {M.~E.}\ \bibnamefont {Gershenson}},\ }\href {\doibase 10.1103/PhysRevLett.112.167001} {\bibfield  {journal} {\bibinfo  {journal} {Phys. Rev. Lett.}\ }\textbf {\bibinfo {volume} {112}},\ \bibinfo {pages} {167001} (\bibinfo {year} {2014})}\BibitemShut {NoStop}%
\bibitem [{\citenamefont {de~Lange}\ \emph {et~al.}(2015)\citenamefont {de~Lange}, \citenamefont {van Heck}, \citenamefont {Bruno}, \citenamefont {van Woerkom}, \citenamefont {Geresdi}, \citenamefont {Plissard}, \citenamefont {Bakkers}, \citenamefont {Akhmerov},\ and\ \citenamefont {DiCarlo}}]{deLange2015Sep}%
  \BibitemOpen
  \bibfield  {author} {\bibinfo {author} {\bibfnamefont {G.}~\bibnamefont {de~Lange}}, \bibinfo {author} {\bibfnamefont {B.}~\bibnamefont {van Heck}}, \bibinfo {author} {\bibfnamefont {A.}~\bibnamefont {Bruno}}, \bibinfo {author} {\bibfnamefont {D.~J.}\ \bibnamefont {van Woerkom}}, \bibinfo {author} {\bibfnamefont {A.}~\bibnamefont {Geresdi}}, \bibinfo {author} {\bibfnamefont {S.~R.}\ \bibnamefont {Plissard}}, \bibinfo {author} {\bibfnamefont {E.~P. A.~M.}\ \bibnamefont {Bakkers}}, \bibinfo {author} {\bibfnamefont {A.~R.}\ \bibnamefont {Akhmerov}}, \ and\ \bibinfo {author} {\bibfnamefont {L.}~\bibnamefont {DiCarlo}},\ }\href {\doibase 10.1103/PhysRevLett.115.127002} {\bibfield  {journal} {\bibinfo  {journal} {Phys. Rev. Lett.}\ }\textbf {\bibinfo {volume} {115}},\ \bibinfo {pages} {127002} (\bibinfo {year} {2015})}\BibitemShut {NoStop}%
\bibitem [{\citenamefont {Smith}\ \emph {et~al.}(2020)\citenamefont {Smith}, \citenamefont {Kou}, \citenamefont {Xiao}, \citenamefont {Vool},\ and\ \citenamefont {Devoret}}]{Smith2020Jan}%
  \BibitemOpen
  \bibfield  {author} {\bibinfo {author} {\bibfnamefont {W.~C.}\ \bibnamefont {Smith}}, \bibinfo {author} {\bibfnamefont {A.}~\bibnamefont {Kou}}, \bibinfo {author} {\bibfnamefont {X.}~\bibnamefont {Xiao}}, \bibinfo {author} {\bibfnamefont {U.}~\bibnamefont {Vool}}, \ and\ \bibinfo {author} {\bibfnamefont {M.~H.}\ \bibnamefont {Devoret}},\ }\href {\doibase 10.1038/s41534-019-0231-2} {\bibfield  {journal} {\bibinfo  {journal} {npj Quantum Inf.}\ }\textbf {\bibinfo {volume} {6}},\ \bibinfo {pages} {1} (\bibinfo {year} {2020})}\BibitemShut {NoStop}%
\bibitem [{\citenamefont {Gyenis}\ \emph {et~al.}(2021)\citenamefont {Gyenis}, \citenamefont {Paolo}, \citenamefont {Koch}, \citenamefont {Blais}, \citenamefont {Houck},\ and\ \citenamefont {Schuster}}]{Gyenis2021Sep}%
  \BibitemOpen
  \bibfield  {author} {\bibinfo {author} {\bibfnamefont {A.}~\bibnamefont {Gyenis}}, \bibinfo {author} {\bibfnamefont {A.~D.}\ \bibnamefont {Paolo}}, \bibinfo {author} {\bibfnamefont {J.}~\bibnamefont {Koch}}, \bibinfo {author} {\bibfnamefont {A.}~\bibnamefont {Blais}}, \bibinfo {author} {\bibfnamefont {A.~A.}\ \bibnamefont {Houck}}, \ and\ \bibinfo {author} {\bibfnamefont {D.~I.}\ \bibnamefont {Schuster}},\ }\href {\doibase 10.1103/prxquantum.2.030101} {\bibfield  {journal} {\bibinfo  {journal} {PRX Quantum}\ }\textbf {\bibinfo {volume} {2}},\ \bibinfo {pages} {030101} (\bibinfo {year} {2021})}\BibitemShut {NoStop}%
\bibitem [{\citenamefont {Schrade}\ \emph {et~al.}(2022)\citenamefont {Schrade}, \citenamefont {Marcus},\ and\ \citenamefont {Gyenis}}]{Schrade2022Jul}%
  \BibitemOpen
  \bibfield  {author} {\bibinfo {author} {\bibfnamefont {C.}~\bibnamefont {Schrade}}, \bibinfo {author} {\bibfnamefont {C.~M.}\ \bibnamefont {Marcus}}, \ and\ \bibinfo {author} {\bibfnamefont {A.}~\bibnamefont {Gyenis}},\ }\href {\doibase 10.1103/PRXQuantum.3.030303} {\bibfield  {journal} {\bibinfo  {journal} {PRX Quantum}\ }\textbf {\bibinfo {volume} {3}},\ \bibinfo {pages} {030303} (\bibinfo {year} {2022})}\BibitemShut {NoStop}%
\bibitem [{\citenamefont {Hover}\ \emph {et~al.}(2012)\citenamefont {Hover}, \citenamefont {Chen}, \citenamefont {Ribeill}, \citenamefont {Zhu}, \citenamefont {Sendelbach},\ and\ \citenamefont {McDermott}}]{Hover2012Feb}%
  \BibitemOpen
  \bibfield  {author} {\bibinfo {author} {\bibfnamefont {D.}~\bibnamefont {Hover}}, \bibinfo {author} {\bibfnamefont {Y.-F.}\ \bibnamefont {Chen}}, \bibinfo {author} {\bibfnamefont {G.~J.}\ \bibnamefont {Ribeill}}, \bibinfo {author} {\bibfnamefont {S.}~\bibnamefont {Zhu}}, \bibinfo {author} {\bibfnamefont {S.}~\bibnamefont {Sendelbach}}, \ and\ \bibinfo {author} {\bibfnamefont {R.}~\bibnamefont {McDermott}},\ }\href {\doibase 10.1063/1.3682309} {\bibfield  {journal} {\bibinfo  {journal} {Appl. Phys. Lett.}\ }\textbf {\bibinfo {volume} {100}},\ \bibinfo {pages} {063503} (\bibinfo {year} {2012})}\BibitemShut {NoStop}%
\bibitem [{\citenamefont {Frattini}\ \emph {et~al.}(2017)\citenamefont {Frattini}, \citenamefont {Vool}, \citenamefont {Shankar}, \citenamefont {Narla}, \citenamefont {Sliwa},\ and\ \citenamefont {Devoret}}]{Frattini2017May}%
  \BibitemOpen
  \bibfield  {author} {\bibinfo {author} {\bibfnamefont {N.~E.}\ \bibnamefont {Frattini}}, \bibinfo {author} {\bibfnamefont {U.}~\bibnamefont {Vool}}, \bibinfo {author} {\bibfnamefont {S.}~\bibnamefont {Shankar}}, \bibinfo {author} {\bibfnamefont {A.}~\bibnamefont {Narla}}, \bibinfo {author} {\bibfnamefont {K.~M.}\ \bibnamefont {Sliwa}}, \ and\ \bibinfo {author} {\bibfnamefont {M.~H.}\ \bibnamefont {Devoret}},\ }\href {\doibase 10.1063/1.4984142} {\bibfield  {journal} {\bibinfo  {journal} {Appl. Phys. Lett.}\ }\textbf {\bibinfo {volume} {110}},\ \bibinfo {pages} {222603} (\bibinfo {year} {2017})}\BibitemShut {NoStop}%
\bibitem [{\citenamefont {Abdo}\ \emph {et~al.}(2013)\citenamefont {Abdo}, \citenamefont {Sliwa}, \citenamefont {Frunzio},\ and\ \citenamefont {Devoret}}]{Abdo2013Jul}%
  \BibitemOpen
  \bibfield  {author} {\bibinfo {author} {\bibfnamefont {B.}~\bibnamefont {Abdo}}, \bibinfo {author} {\bibfnamefont {K.}~\bibnamefont {Sliwa}}, \bibinfo {author} {\bibfnamefont {L.}~\bibnamefont {Frunzio}}, \ and\ \bibinfo {author} {\bibfnamefont {M.}~\bibnamefont {Devoret}},\ }\href {\doibase 10.1103/PhysRevX.3.031001} {\bibfield  {journal} {\bibinfo  {journal} {Phys. Rev. X}\ }\textbf {\bibinfo {volume} {3}},\ \bibinfo {pages} {031001} (\bibinfo {year} {2013})}\BibitemShut {NoStop}%
\bibitem [{\citenamefont {Sivak}\ \emph {et~al.}(2020)\citenamefont {Sivak}, \citenamefont {Shankar}, \citenamefont {Liu}, \citenamefont {Aumentado},\ and\ \citenamefont {Devoret}}]{Sivak2020Feb}%
  \BibitemOpen
  \bibfield  {author} {\bibinfo {author} {\bibfnamefont {V.~V.}\ \bibnamefont {Sivak}}, \bibinfo {author} {\bibfnamefont {S.}~\bibnamefont {Shankar}}, \bibinfo {author} {\bibfnamefont {G.}~\bibnamefont {Liu}}, \bibinfo {author} {\bibfnamefont {J.}~\bibnamefont {Aumentado}}, \ and\ \bibinfo {author} {\bibfnamefont {M.~H.}\ \bibnamefont {Devoret}},\ }\href {\doibase 10.1103/PhysRevApplied.13.024014} {\bibfield  {journal} {\bibinfo  {journal} {Phys. Rev. Appl.}\ }\textbf {\bibinfo {volume} {13}},\ \bibinfo {pages} {024014} (\bibinfo {year} {2020})}\BibitemShut {NoStop}%
\bibitem [{\citenamefont {Schrade}\ and\ \citenamefont {Fatemi}(2023)}]{Schrade2023Oct}%
  \BibitemOpen
  \bibfield  {author} {\bibinfo {author} {\bibfnamefont {C.}~\bibnamefont {Schrade}}\ and\ \bibinfo {author} {\bibfnamefont {V.}~\bibnamefont {Fatemi}},\ }\href {\doibase 10.48550/arXiv.2310.12198} {\bibfield  {journal} {\bibinfo  {journal} {ArXiv}\ ,\ \bibinfo {pages} {2310.12198}} (\bibinfo {year} {2023})}\BibitemShut {NoStop}%
\bibitem [{\citenamefont {Baumgartner}\ \emph {et~al.}(2022)\citenamefont {Baumgartner}, \citenamefont {Fuchs}, \citenamefont {Costa}, \citenamefont {Reinhardt}, \citenamefont {Gronin}, \citenamefont {Gardner}, \citenamefont {Lindemann}, \citenamefont {Manfra}, \citenamefont {Faria~Junior}, \citenamefont {Kochan}, \citenamefont {Fabian}, \citenamefont {Paradiso},\ and\ \citenamefont {Strunk}}]{Baumgartner2022Jan}%
  \BibitemOpen
  \bibfield  {author} {\bibinfo {author} {\bibfnamefont {C.}~\bibnamefont {Baumgartner}}, \bibinfo {author} {\bibfnamefont {L.}~\bibnamefont {Fuchs}}, \bibinfo {author} {\bibfnamefont {A.}~\bibnamefont {Costa}}, \bibinfo {author} {\bibfnamefont {S.}~\bibnamefont {Reinhardt}}, \bibinfo {author} {\bibfnamefont {S.}~\bibnamefont {Gronin}}, \bibinfo {author} {\bibfnamefont {G.~C.}\ \bibnamefont {Gardner}}, \bibinfo {author} {\bibfnamefont {T.}~\bibnamefont {Lindemann}}, \bibinfo {author} {\bibfnamefont {M.~J.}\ \bibnamefont {Manfra}}, \bibinfo {author} {\bibfnamefont {P.~E.}\ \bibnamefont {Faria~Junior}}, \bibinfo {author} {\bibfnamefont {D.}~\bibnamefont {Kochan}}, \bibinfo {author} {\bibfnamefont {J.}~\bibnamefont {Fabian}}, \bibinfo {author} {\bibfnamefont {N.}~\bibnamefont {Paradiso}}, \ and\ \bibinfo {author} {\bibfnamefont {C.}~\bibnamefont {Strunk}},\ }\href {\doibase 10.1038/s41565-021-01009-9} {\bibfield  {journal} {\bibinfo  {journal} {Nat. Nanotechnol.}\ }\textbf {\bibinfo {volume} {17}},\ \bibinfo
  {pages} {39} (\bibinfo {year} {2022})}\BibitemShut {NoStop}%
\bibitem [{\citenamefont {Souto}\ \emph {et~al.}(2022)\citenamefont {Souto}, \citenamefont {Leijnse},\ and\ \citenamefont {Schrade}}]{Souto2022May}%
  \BibitemOpen
  \bibfield  {author} {\bibinfo {author} {\bibfnamefont {R.~S.}\ \bibnamefont {Souto}}, \bibinfo {author} {\bibfnamefont {M.}~\bibnamefont {Leijnse}}, \ and\ \bibinfo {author} {\bibfnamefont {C.}~\bibnamefont {Schrade}},\ }\href {\doibase 10.1103/PhysRevLett.129.267702} {\bibfield  {journal} {\bibinfo  {journal} {Phys. Rev. Lett.}\ }\textbf {\bibinfo {volume} {129}},\ \bibinfo {pages} {267702} (\bibinfo {year} {2022})}\BibitemShut {NoStop}%
\bibitem [{\citenamefont {Ciaccia}\ \emph {et~al.}(2023)\citenamefont {Ciaccia}, \citenamefont {Haller}, \citenamefont {Drachmann}, \citenamefont {Lindemann}, \citenamefont {Manfra}, \citenamefont {Schrade},\ and\ \citenamefont {Sch{\ifmmode\ddot{o}\else\"{o}\fi}nenberger}}]{Ciaccia2023Apr}%
  \BibitemOpen
  \bibfield  {author} {\bibinfo {author} {\bibfnamefont {C.}~\bibnamefont {Ciaccia}}, \bibinfo {author} {\bibfnamefont {R.}~\bibnamefont {Haller}}, \bibinfo {author} {\bibfnamefont {A.~C.~C.}\ \bibnamefont {Drachmann}}, \bibinfo {author} {\bibfnamefont {T.}~\bibnamefont {Lindemann}}, \bibinfo {author} {\bibfnamefont {M.~J.}\ \bibnamefont {Manfra}}, \bibinfo {author} {\bibfnamefont {C.}~\bibnamefont {Schrade}}, \ and\ \bibinfo {author} {\bibfnamefont {C.}~\bibnamefont {Sch{\ifmmode\ddot{o}\else\"{o}\fi}nenberger}},\ }\href {\doibase 10.1103/PhysRevResearch.5.033131} {\bibfield  {journal} {\bibinfo  {journal} {Phys. Rev. Res.}\ }\textbf {\bibinfo {volume} {5}},\ \bibinfo {pages} {033131} (\bibinfo {year} {2023})}\BibitemShut {NoStop}%
\bibitem [{\citenamefont {Bozkurt}\ \emph {et~al.}(2023)\citenamefont {Bozkurt}, \citenamefont {Brookman}, \citenamefont {Fatemi},\ and\ \citenamefont {Akhmerov}}]{Bozkurt2023Jul}%
  \BibitemOpen
  \bibfield  {author} {\bibinfo {author} {\bibfnamefont {A.~M.}\ \bibnamefont {Bozkurt}}, \bibinfo {author} {\bibfnamefont {J.}~\bibnamefont {Brookman}}, \bibinfo {author} {\bibfnamefont {V.}~\bibnamefont {Fatemi}}, \ and\ \bibinfo {author} {\bibfnamefont {A.~R.}\ \bibnamefont {Akhmerov}},\ }\href {\doibase 10.21468/SciPostPhys.15.5.204} {\bibfield  {journal} {\bibinfo  {journal} {SciPost Phys.}\ }\textbf {\bibinfo {volume} {15}},\ \bibinfo {pages} {204} (\bibinfo {year} {2023})}\BibitemShut {NoStop}%
\bibitem [{\citenamefont {Barash}(2018)}]{Barash2018Jun}%
  \BibitemOpen
  \bibfield  {author} {\bibinfo {author} {\bibfnamefont {{\relax Yu}.~S.}\ \bibnamefont {Barash}},\ }\href {\doibase 10.1103/PhysRevB.97.224509} {\bibfield  {journal} {\bibinfo  {journal} {Phys. Rev. B}\ }\textbf {\bibinfo {volume} {97}},\ \bibinfo {pages} {224509} (\bibinfo {year} {2018})}\BibitemShut {NoStop}%
\bibitem [{\citenamefont {Coraiola}\ \emph {et~al.}(2023)\citenamefont {Coraiola}, \citenamefont {Haxell}, \citenamefont {Sabonis}, \citenamefont {Weisbrich}, \citenamefont {Svetogorov}, \citenamefont {Hinderling}, \citenamefont {Kate}, \citenamefont {Cheah}, \citenamefont {Krizek}, \citenamefont {Schott}, \citenamefont {Wegscheider}, \citenamefont {Cuevas}, \citenamefont {Belzig},\ and\ \citenamefont {Nichele}}]{Coraiola2023May}%
  \BibitemOpen
  \bibfield  {author} {\bibinfo {author} {\bibfnamefont {M.}~\bibnamefont {Coraiola}}, \bibinfo {author} {\bibfnamefont {D.~Z.}\ \bibnamefont {Haxell}}, \bibinfo {author} {\bibfnamefont {D.}~\bibnamefont {Sabonis}}, \bibinfo {author} {\bibfnamefont {H.}~\bibnamefont {Weisbrich}}, \bibinfo {author} {\bibfnamefont {A.~E.}\ \bibnamefont {Svetogorov}}, \bibinfo {author} {\bibfnamefont {M.}~\bibnamefont {Hinderling}}, \bibinfo {author} {\bibfnamefont {S.~C.~t.}\ \bibnamefont {Kate}}, \bibinfo {author} {\bibfnamefont {E.}~\bibnamefont {Cheah}}, \bibinfo {author} {\bibfnamefont {F.}~\bibnamefont {Krizek}}, \bibinfo {author} {\bibfnamefont {R.}~\bibnamefont {Schott}}, \bibinfo {author} {\bibfnamefont {W.}~\bibnamefont {Wegscheider}}, \bibinfo {author} {\bibfnamefont {J.~C.}\ \bibnamefont {Cuevas}}, \bibinfo {author} {\bibfnamefont {W.}~\bibnamefont {Belzig}}, \ and\ \bibinfo {author} {\bibfnamefont {F.}~\bibnamefont {Nichele}},\ }\href {\doibase 10.1038/s41467-023-42356-6} {\bibfield  {journal} {\bibinfo  {journal}
  {Nat. Commun.}\ }\textbf {\bibinfo {volume} {14}},\ \bibinfo {pages} {6784} (\bibinfo {year} {2023})}\BibitemShut {NoStop}%
\bibitem [{\citenamefont {Matsuo}\ \emph {et~al.}(2023)\citenamefont {Matsuo}, \citenamefont {Imoto}, \citenamefont {Yokoyama}, \citenamefont {Sato}, \citenamefont {Lindemann}, \citenamefont {Gronin}, \citenamefont {Gardner}, \citenamefont {Manfra},\ and\ \citenamefont {Tarucha}}]{Matsuo2023Dec}%
  \BibitemOpen
  \bibfield  {author} {\bibinfo {author} {\bibfnamefont {S.}~\bibnamefont {Matsuo}}, \bibinfo {author} {\bibfnamefont {T.}~\bibnamefont {Imoto}}, \bibinfo {author} {\bibfnamefont {T.}~\bibnamefont {Yokoyama}}, \bibinfo {author} {\bibfnamefont {Y.}~\bibnamefont {Sato}}, \bibinfo {author} {\bibfnamefont {T.}~\bibnamefont {Lindemann}}, \bibinfo {author} {\bibfnamefont {S.}~\bibnamefont {Gronin}}, \bibinfo {author} {\bibfnamefont {G.~C.}\ \bibnamefont {Gardner}}, \bibinfo {author} {\bibfnamefont {M.~J.}\ \bibnamefont {Manfra}}, \ and\ \bibinfo {author} {\bibfnamefont {S.}~\bibnamefont {Tarucha}},\ }\href {\doibase 10.1126/sciadv.adj3698} {\bibfield  {journal} {\bibinfo  {journal} {Sci. Adv.}\ }\textbf {\bibinfo {volume} {9}},\ \bibinfo {pages} {adj3698} (\bibinfo {year} {2023})}\BibitemShut {NoStop}%
\bibitem [{\citenamefont {Dou{\ifmmode\mbox{\c{c}}\else\c{c}\fi}ot}\ and\ \citenamefont {Vidal}(2002)}]{Doucot2002May}%
  \BibitemOpen
  \bibfield  {author} {\bibinfo {author} {\bibfnamefont {B.}~\bibnamefont {Dou{\ifmmode\mbox{\c{c}}\else\c{c}\fi}ot}}\ and\ \bibinfo {author} {\bibfnamefont {J.}~\bibnamefont {Vidal}},\ }\href {\doibase 10.1103/PhysRevLett.88.227005} {\bibfield  {journal} {\bibinfo  {journal} {Phys. Rev. Lett.}\ }\textbf {\bibinfo {volume} {88}},\ \bibinfo {pages} {227005} (\bibinfo {year} {2002})}\BibitemShut {NoStop}%
\bibitem [{\citenamefont {Pop}\ \emph {et~al.}(2008)\citenamefont {Pop}, \citenamefont {Hasselbach}, \citenamefont {Buisson}, \citenamefont {Guichard}, \citenamefont {Pannetier},\ and\ \citenamefont {Protopopov}}]{Pop2008Sep}%
  \BibitemOpen
  \bibfield  {author} {\bibinfo {author} {\bibfnamefont {I.~M.}\ \bibnamefont {Pop}}, \bibinfo {author} {\bibfnamefont {K.}~\bibnamefont {Hasselbach}}, \bibinfo {author} {\bibfnamefont {O.}~\bibnamefont {Buisson}}, \bibinfo {author} {\bibfnamefont {W.}~\bibnamefont {Guichard}}, \bibinfo {author} {\bibfnamefont {B.}~\bibnamefont {Pannetier}}, \ and\ \bibinfo {author} {\bibfnamefont {I.}~\bibnamefont {Protopopov}},\ }\href {\doibase 10.1103/PhysRevB.78.104504} {\bibfield  {journal} {\bibinfo  {journal} {Phys. Rev. B}\ }\textbf {\bibinfo {volume} {78}},\ \bibinfo {pages} {104504} (\bibinfo {year} {2008})}\BibitemShut {NoStop}%
\bibitem [{\citenamefont {Gladchenko}\ \emph {et~al.}(2009)\citenamefont {Gladchenko}, \citenamefont {Olaya}, \citenamefont {Dupont-Ferrier}, \citenamefont {Dou{\ifmmode\mbox{\c{c}}\else\c{c}\fi}ot}, \citenamefont {Ioffe},\ and\ \citenamefont {Gershenson}}]{Gladchenko2009Jan}%
  \BibitemOpen
  \bibfield  {author} {\bibinfo {author} {\bibfnamefont {S.}~\bibnamefont {Gladchenko}}, \bibinfo {author} {\bibfnamefont {D.}~\bibnamefont {Olaya}}, \bibinfo {author} {\bibfnamefont {E.}~\bibnamefont {Dupont-Ferrier}}, \bibinfo {author} {\bibfnamefont {B.}~\bibnamefont {Dou{\ifmmode\mbox{\c{c}}\else\c{c}\fi}ot}}, \bibinfo {author} {\bibfnamefont {L.~B.}\ \bibnamefont {Ioffe}}, \ and\ \bibinfo {author} {\bibfnamefont {M.~E.}\ \bibnamefont {Gershenson}},\ }\href {\doibase 10.1038/nphys1151} {\bibfield  {journal} {\bibinfo  {journal} {Nat. Phys.}\ }\textbf {\bibinfo {volume} {5}},\ \bibinfo {pages} {48} (\bibinfo {year} {2009})}\BibitemShut {NoStop}%
\bibitem [{\citenamefont {Beenakker}(1991)}]{Beenakker1991Dec}%
  \BibitemOpen
  \bibfield  {author} {\bibinfo {author} {\bibfnamefont {C.~W.~J.}\ \bibnamefont {Beenakker}},\ }\href {\doibase 10.1103/PhysRevLett.67.3836} {\bibfield  {journal} {\bibinfo  {journal} {Phys. Rev. Lett.}\ }\textbf {\bibinfo {volume} {67}},\ \bibinfo {pages} {3836} (\bibinfo {year} {1991})}\BibitemShut {NoStop}%
\bibitem [{\citenamefont {Woerkom}\ \emph {et~al.}(2017)\citenamefont {Woerkom}, \citenamefont {Proutski}, \citenamefont {Heck}, \citenamefont {Bouman}, \citenamefont {Väyrynen}, \citenamefont {Glazman}, \citenamefont {Krogstrup}, \citenamefont {Nygård}, \citenamefont {Kouwenhoven},\ and\ \citenamefont {Geresdi}}]{vanWoerkom2017Sep}%
  \BibitemOpen
  \bibfield  {author} {\bibinfo {author} {\bibfnamefont {D.~J.~v.}\ \bibnamefont {Woerkom}}, \bibinfo {author} {\bibfnamefont {A.}~\bibnamefont {Proutski}}, \bibinfo {author} {\bibfnamefont {B.~v.}\ \bibnamefont {Heck}}, \bibinfo {author} {\bibfnamefont {D.}~\bibnamefont {Bouman}}, \bibinfo {author} {\bibfnamefont {J.~I.}\ \bibnamefont {Väyrynen}}, \bibinfo {author} {\bibfnamefont {L.~I.}\ \bibnamefont {Glazman}}, \bibinfo {author} {\bibfnamefont {P.}~\bibnamefont {Krogstrup}}, \bibinfo {author} {\bibfnamefont {J.}~\bibnamefont {Nygård}}, \bibinfo {author} {\bibfnamefont {L.~P.}\ \bibnamefont {Kouwenhoven}}, \ and\ \bibinfo {author} {\bibfnamefont {A.}~\bibnamefont {Geresdi}},\ }\href {\doibase 10.1038/nphys4150} {\bibfield  {journal} {\bibinfo  {journal} {Nat. Phys.}\ }\textbf {\bibinfo {volume} {13}},\ \bibinfo {pages} {876} (\bibinfo {year} {2017})}\BibitemShut {NoStop}%
\bibitem [{\citenamefont {Spanton}\ \emph {et~al.}(2017)\citenamefont {Spanton}, \citenamefont {Deng}, \citenamefont {Vaitiek{\ifmmode\dot{e}\else\.{e}\fi}nas}, \citenamefont {Krogstrup}, \citenamefont {Nyg{\aa}rd}, \citenamefont {Marcus},\ and\ \citenamefont {Moler}}]{Spanton2017Dec}%
  \BibitemOpen
  \bibfield  {author} {\bibinfo {author} {\bibfnamefont {E.~M.}\ \bibnamefont {Spanton}}, \bibinfo {author} {\bibfnamefont {M.}~\bibnamefont {Deng}}, \bibinfo {author} {\bibfnamefont {S.}~\bibnamefont {Vaitiek{\ifmmode\dot{e}\else\.{e}\fi}nas}}, \bibinfo {author} {\bibfnamefont {P.}~\bibnamefont {Krogstrup}}, \bibinfo {author} {\bibfnamefont {J.}~\bibnamefont {Nyg{\aa}rd}}, \bibinfo {author} {\bibfnamefont {C.~M.}\ \bibnamefont {Marcus}}, \ and\ \bibinfo {author} {\bibfnamefont {K.~A.}\ \bibnamefont {Moler}},\ }\href {\doibase 10.1038/nphys4224} {\bibfield  {journal} {\bibinfo  {journal} {Nat. Phys.}\ }\textbf {\bibinfo {volume} {13}},\ \bibinfo {pages} {1177} (\bibinfo {year} {2017})}\BibitemShut {NoStop}%
\bibitem [{\citenamefont {Larsen}\ \emph {et~al.}(2020)\citenamefont {Larsen}, \citenamefont {Gershenson}, \citenamefont {Casparis}, \citenamefont {Kringh{\o}j}, \citenamefont {Pearson}, \citenamefont {McNeil}, \citenamefont {Kuemmeth}, \citenamefont {Krogstrup}, \citenamefont {Petersson},\ and\ \citenamefont {Marcus}}]{Larsen2020Jul}%
  \BibitemOpen
  \bibfield  {author} {\bibinfo {author} {\bibfnamefont {T.~W.}\ \bibnamefont {Larsen}}, \bibinfo {author} {\bibfnamefont {M.~E.}\ \bibnamefont {Gershenson}}, \bibinfo {author} {\bibfnamefont {L.}~\bibnamefont {Casparis}}, \bibinfo {author} {\bibfnamefont {A.}~\bibnamefont {Kringh{\o}j}}, \bibinfo {author} {\bibfnamefont {N.~J.}\ \bibnamefont {Pearson}}, \bibinfo {author} {\bibfnamefont {R.~P.~G.}\ \bibnamefont {McNeil}}, \bibinfo {author} {\bibfnamefont {F.}~\bibnamefont {Kuemmeth}}, \bibinfo {author} {\bibfnamefont {P.}~\bibnamefont {Krogstrup}}, \bibinfo {author} {\bibfnamefont {K.~D.}\ \bibnamefont {Petersson}}, \ and\ \bibinfo {author} {\bibfnamefont {C.~M.}\ \bibnamefont {Marcus}},\ }\href {\doibase 10.1103/PhysRevLett.125.056801} {\bibfield  {journal} {\bibinfo  {journal} {Phys. Rev. Lett.}\ }\textbf {\bibinfo {volume} {125}},\ \bibinfo {pages} {056801} (\bibinfo {year} {2020})}\BibitemShut {NoStop}%
\bibitem [{\citenamefont {Ueda}\ \emph {et~al.}(2020)\citenamefont {Ueda}, \citenamefont {Matsuo}, \citenamefont {Kamata}, \citenamefont {Sato}, \citenamefont {Takeshige}, \citenamefont {Li}, \citenamefont {Samuelson}, \citenamefont {Xu},\ and\ \citenamefont {Tarucha}}]{Ueda2020Sep}%
  \BibitemOpen
  \bibfield  {author} {\bibinfo {author} {\bibfnamefont {K.}~\bibnamefont {Ueda}}, \bibinfo {author} {\bibfnamefont {S.}~\bibnamefont {Matsuo}}, \bibinfo {author} {\bibfnamefont {H.}~\bibnamefont {Kamata}}, \bibinfo {author} {\bibfnamefont {Y.}~\bibnamefont {Sato}}, \bibinfo {author} {\bibfnamefont {Y.}~\bibnamefont {Takeshige}}, \bibinfo {author} {\bibfnamefont {K.}~\bibnamefont {Li}}, \bibinfo {author} {\bibfnamefont {L.}~\bibnamefont {Samuelson}}, \bibinfo {author} {\bibfnamefont {H.}~\bibnamefont {Xu}}, \ and\ \bibinfo {author} {\bibfnamefont {S.}~\bibnamefont {Tarucha}},\ }\href {\doibase 10.1103/PhysRevResearch.2.033435} {\bibfield  {journal} {\bibinfo  {journal} {Phys. Rev. Res.}\ }\textbf {\bibinfo {volume} {2}},\ \bibinfo {pages} {033435} (\bibinfo {year} {2020})}\BibitemShut {NoStop}%
\bibitem [{\citenamefont {Nichele}\ \emph {et~al.}(2020)\citenamefont {Nichele}, \citenamefont {Portol{\ifmmode\acute{e}\else\'{e}\fi}s}, \citenamefont {Fornieri}, \citenamefont {Whiticar}, \citenamefont {Drachmann}, \citenamefont {Gronin}, \citenamefont {Wang}, \citenamefont {Gardner}, \citenamefont {Thomas}, \citenamefont {Hatke}, \citenamefont {Manfra},\ and\ \citenamefont {Marcus}}]{Nichele2020Jun}%
  \BibitemOpen
  \bibfield  {author} {\bibinfo {author} {\bibfnamefont {F.}~\bibnamefont {Nichele}}, \bibinfo {author} {\bibfnamefont {E.}~\bibnamefont {Portol{\ifmmode\acute{e}\else\'{e}\fi}s}}, \bibinfo {author} {\bibfnamefont {A.}~\bibnamefont {Fornieri}}, \bibinfo {author} {\bibfnamefont {A.~M.}\ \bibnamefont {Whiticar}}, \bibinfo {author} {\bibfnamefont {A.~C.~C.}\ \bibnamefont {Drachmann}}, \bibinfo {author} {\bibfnamefont {S.}~\bibnamefont {Gronin}}, \bibinfo {author} {\bibfnamefont {T.}~\bibnamefont {Wang}}, \bibinfo {author} {\bibfnamefont {G.~C.}\ \bibnamefont {Gardner}}, \bibinfo {author} {\bibfnamefont {C.}~\bibnamefont {Thomas}}, \bibinfo {author} {\bibfnamefont {A.~T.}\ \bibnamefont {Hatke}}, \bibinfo {author} {\bibfnamefont {M.~J.}\ \bibnamefont {Manfra}}, \ and\ \bibinfo {author} {\bibfnamefont {C.~M.}\ \bibnamefont {Marcus}},\ }\href {\doibase 10.1103/PhysRevLett.124.226801} {\bibfield  {journal} {\bibinfo  {journal} {Phys. Rev. Lett.}\ }\textbf {\bibinfo {volume} {124}},\ \bibinfo {pages} {226801}
  (\bibinfo {year} {2020})}\BibitemShut {NoStop}%
\bibitem [{\citenamefont {Ciaccia}\ \emph {et~al.}(2024)\citenamefont {Ciaccia}, \citenamefont {Haller}, \citenamefont {Drachmann}, \citenamefont {Lindemann}, \citenamefont {Manfra}, \citenamefont {Schrade},\ and\ \citenamefont {Schönenberger}}]{Ciaccia2023Jun}%
  \BibitemOpen
  \bibfield  {author} {\bibinfo {author} {\bibfnamefont {C.}~\bibnamefont {Ciaccia}}, \bibinfo {author} {\bibfnamefont {R.}~\bibnamefont {Haller}}, \bibinfo {author} {\bibfnamefont {A.~C.~C.}\ \bibnamefont {Drachmann}}, \bibinfo {author} {\bibfnamefont {T.}~\bibnamefont {Lindemann}}, \bibinfo {author} {\bibfnamefont {M.~J.}\ \bibnamefont {Manfra}}, \bibinfo {author} {\bibfnamefont {C.}~\bibnamefont {Schrade}}, \ and\ \bibinfo {author} {\bibfnamefont {C.}~\bibnamefont {Schönenberger}},\ }\href {\doibase 10.1038/s42005-024-01531-x} {\bibfield  {journal} {\bibinfo  {journal} {Commun. Phys.}\ }\textbf {\bibinfo {volume} {7}},\ \bibinfo {pages} {41} (\bibinfo {year} {2024})}\BibitemShut {NoStop}%
\bibitem [{\citenamefont {Valentini}\ \emph {et~al.}(2024)\citenamefont {Valentini}, \citenamefont {Sagi}, \citenamefont {Baghumyan}, \citenamefont {Gijsel}, \citenamefont {Jung}, \citenamefont {Calcaterra}, \citenamefont {Ballabio}, \citenamefont {Servin}, \citenamefont {Aggarwal}, \citenamefont {Janik}, \citenamefont {Adletzberger}, \citenamefont {Souto}, \citenamefont {Leijnse}, \citenamefont {Danon}, \citenamefont {Schrade}, \citenamefont {Bakkers}, \citenamefont {Chrastina}, \citenamefont {Isella},\ and\ \citenamefont {Katsaros}}]{Valentini2023Jun}%
  \BibitemOpen
  \bibfield  {author} {\bibinfo {author} {\bibfnamefont {M.}~\bibnamefont {Valentini}}, \bibinfo {author} {\bibfnamefont {O.}~\bibnamefont {Sagi}}, \bibinfo {author} {\bibfnamefont {L.}~\bibnamefont {Baghumyan}}, \bibinfo {author} {\bibfnamefont {T.~d.}\ \bibnamefont {Gijsel}}, \bibinfo {author} {\bibfnamefont {J.}~\bibnamefont {Jung}}, \bibinfo {author} {\bibfnamefont {S.}~\bibnamefont {Calcaterra}}, \bibinfo {author} {\bibfnamefont {A.}~\bibnamefont {Ballabio}}, \bibinfo {author} {\bibfnamefont {J.~A.}\ \bibnamefont {Servin}}, \bibinfo {author} {\bibfnamefont {K.}~\bibnamefont {Aggarwal}}, \bibinfo {author} {\bibfnamefont {M.}~\bibnamefont {Janik}}, \bibinfo {author} {\bibfnamefont {T.}~\bibnamefont {Adletzberger}}, \bibinfo {author} {\bibfnamefont {R.~S.}\ \bibnamefont {Souto}}, \bibinfo {author} {\bibfnamefont {M.}~\bibnamefont {Leijnse}}, \bibinfo {author} {\bibfnamefont {J.}~\bibnamefont {Danon}}, \bibinfo {author} {\bibfnamefont {C.}~\bibnamefont {Schrade}}, \bibinfo {author} {\bibfnamefont
  {E.}~\bibnamefont {Bakkers}}, \bibinfo {author} {\bibfnamefont {D.}~\bibnamefont {Chrastina}}, \bibinfo {author} {\bibfnamefont {G.}~\bibnamefont {Isella}}, \ and\ \bibinfo {author} {\bibfnamefont {G.}~\bibnamefont {Katsaros}},\ }\href {\doibase 10.1038/s41467-023-44114-0} {\bibfield  {journal} {\bibinfo  {journal} {Nat. Commun.}\ }\textbf {\bibinfo {volume} {15}},\ \bibinfo {pages} {169} (\bibinfo {year} {2024})}\BibitemShut {NoStop}%
\bibitem [{\citenamefont {Leblanc}\ \emph {et~al.}(2023)\citenamefont {Leblanc}, \citenamefont {Tangchingchai}, \citenamefont {Momtaz}, \citenamefont {Kiyooka}, \citenamefont {Hartmann}, \citenamefont {Fernandez-Bada}, \citenamefont {Brun-Barriere}, \citenamefont {Schmitt}, \citenamefont {Zihlmann}, \citenamefont {Maurand}, \citenamefont {Dumur}, \citenamefont {De~Franceschi},\ and\ \citenamefont {Lefloch}}]{Leblanc2023Nov}%
  \BibitemOpen
  \bibfield  {author} {\bibinfo {author} {\bibfnamefont {A.}~\bibnamefont {Leblanc}}, \bibinfo {author} {\bibfnamefont {C.}~\bibnamefont {Tangchingchai}}, \bibinfo {author} {\bibfnamefont {Z.~S.}\ \bibnamefont {Momtaz}}, \bibinfo {author} {\bibfnamefont {E.}~\bibnamefont {Kiyooka}}, \bibinfo {author} {\bibfnamefont {J.-M.}\ \bibnamefont {Hartmann}}, \bibinfo {author} {\bibfnamefont {G.~T.}\ \bibnamefont {Fernandez-Bada}}, \bibinfo {author} {\bibfnamefont {B.}~\bibnamefont {Brun-Barriere}}, \bibinfo {author} {\bibfnamefont {V.}~\bibnamefont {Schmitt}}, \bibinfo {author} {\bibfnamefont {S.}~\bibnamefont {Zihlmann}}, \bibinfo {author} {\bibfnamefont {R.}~\bibnamefont {Maurand}}, \bibinfo {author} {\bibfnamefont {{\ifmmode\acute{E}\else\'{E}\fi}.}~\bibnamefont {Dumur}}, \bibinfo {author} {\bibfnamefont {S.}~\bibnamefont {De~Franceschi}}, \ and\ \bibinfo {author} {\bibfnamefont {F.}~\bibnamefont {Lefloch}},\ }\href {\doibase 10.48550/arXiv.2311.15371} {\bibfield  {journal} {\bibinfo  {journal} {arXiv}\ ,\ \bibinfo
  {pages} {2311.15371}} (\bibinfo {year} {2023})}\BibitemShut {NoStop}%
\bibitem [{\citenamefont {English}\ \emph {et~al.}(2016)\citenamefont {English}, \citenamefont {Hamilton}, \citenamefont {Chialvo}, \citenamefont {Moraru}, \citenamefont {Mason},\ and\ \citenamefont {Van~Harlingen}}]{English2016Sep}%
  \BibitemOpen
  \bibfield  {author} {\bibinfo {author} {\bibfnamefont {C.~D.}\ \bibnamefont {English}}, \bibinfo {author} {\bibfnamefont {D.~R.}\ \bibnamefont {Hamilton}}, \bibinfo {author} {\bibfnamefont {C.}~\bibnamefont {Chialvo}}, \bibinfo {author} {\bibfnamefont {I.~C.}\ \bibnamefont {Moraru}}, \bibinfo {author} {\bibfnamefont {N.}~\bibnamefont {Mason}}, \ and\ \bibinfo {author} {\bibfnamefont {D.~J.}\ \bibnamefont {Van~Harlingen}},\ }\href {\doibase 10.1103/PhysRevB.94.115435} {\bibfield  {journal} {\bibinfo  {journal} {Phys. Rev. B}\ }\textbf {\bibinfo {volume} {94}},\ \bibinfo {pages} {115435} (\bibinfo {year} {2016})}\BibitemShut {NoStop}%
\bibitem [{\citenamefont {Nanda}\ \emph {et~al.}(2017)\citenamefont {Nanda}, \citenamefont {Aguilera-Servin}, \citenamefont {Rakyta}, \citenamefont {Kormányos}, \citenamefont {Kleiner}, \citenamefont {Koelle}, \citenamefont {Watanabe}, \citenamefont {Taniguchi}, \citenamefont {Vandersypen},\ and\ \citenamefont {Goswami}}]{Nanda2017}%
  \BibitemOpen
  \bibfield  {author} {\bibinfo {author} {\bibfnamefont {G.}~\bibnamefont {Nanda}}, \bibinfo {author} {\bibfnamefont {J.~L.}\ \bibnamefont {Aguilera-Servin}}, \bibinfo {author} {\bibfnamefont {P.}~\bibnamefont {Rakyta}}, \bibinfo {author} {\bibfnamefont {A.}~\bibnamefont {Kormányos}}, \bibinfo {author} {\bibfnamefont {R.}~\bibnamefont {Kleiner}}, \bibinfo {author} {\bibfnamefont {D.}~\bibnamefont {Koelle}}, \bibinfo {author} {\bibfnamefont {K.}~\bibnamefont {Watanabe}}, \bibinfo {author} {\bibfnamefont {T.}~\bibnamefont {Taniguchi}}, \bibinfo {author} {\bibfnamefont {L.~M.~K.}\ \bibnamefont {Vandersypen}}, \ and\ \bibinfo {author} {\bibfnamefont {S.}~\bibnamefont {Goswami}},\ }\href {\doibase 10.1021/acs.nanolett.7b00097} {\bibfield  {journal} {\bibinfo  {journal} {Nano Lett.}\ }\textbf {\bibinfo {volume} {17}},\ \bibinfo {pages} {3396} (\bibinfo {year} {2017})}\BibitemShut {NoStop}%
\bibitem [{\citenamefont {Portol{\ifmmode\acute{e}\else\'{e}\fi}s}\ \emph {et~al.}(2022)\citenamefont {Portol{\ifmmode\acute{e}\else\'{e}\fi}s}, \citenamefont {Iwakiri}, \citenamefont {Zheng}, \citenamefont {Rickhaus}, \citenamefont {Taniguchi}, \citenamefont {Watanabe}, \citenamefont {Ihn}, \citenamefont {Ensslin},\ and\ \citenamefont {de~Vries}}]{Portoles2022Nov}%
  \BibitemOpen
  \bibfield  {author} {\bibinfo {author} {\bibfnamefont {E.}~\bibnamefont {Portol{\ifmmode\acute{e}\else\'{e}\fi}s}}, \bibinfo {author} {\bibfnamefont {S.}~\bibnamefont {Iwakiri}}, \bibinfo {author} {\bibfnamefont {G.}~\bibnamefont {Zheng}}, \bibinfo {author} {\bibfnamefont {P.}~\bibnamefont {Rickhaus}}, \bibinfo {author} {\bibfnamefont {T.}~\bibnamefont {Taniguchi}}, \bibinfo {author} {\bibfnamefont {K.}~\bibnamefont {Watanabe}}, \bibinfo {author} {\bibfnamefont {T.}~\bibnamefont {Ihn}}, \bibinfo {author} {\bibfnamefont {K.}~\bibnamefont {Ensslin}}, \ and\ \bibinfo {author} {\bibfnamefont {F.~K.}\ \bibnamefont {de~Vries}},\ }\href {\doibase 10.1038/s41565-022-01222-0} {\bibfield  {journal} {\bibinfo  {journal} {Nat. Nanotechnol.}\ }\textbf {\bibinfo {volume} {17}},\ \bibinfo {pages} {1159} (\bibinfo {year} {2022})}\BibitemShut {NoStop}%
\bibitem [{\citenamefont {Sochnikov}\ \emph {et~al.}(2015)\citenamefont {Sochnikov}, \citenamefont {Maier}, \citenamefont {Watson}, \citenamefont {Kirtley}, \citenamefont {Gould}, \citenamefont {Tkachov}, \citenamefont {Hankiewicz}, \citenamefont {Br{\ifmmode\ddot{u}\else\"{u}\fi}ne}, \citenamefont {Buhmann}, \citenamefont {Molenkamp},\ and\ \citenamefont {Moler}}]{Sochnikov2015Feb}%
  \BibitemOpen
  \bibfield  {author} {\bibinfo {author} {\bibfnamefont {I.}~\bibnamefont {Sochnikov}}, \bibinfo {author} {\bibfnamefont {L.}~\bibnamefont {Maier}}, \bibinfo {author} {\bibfnamefont {C.~A.}\ \bibnamefont {Watson}}, \bibinfo {author} {\bibfnamefont {J.~R.}\ \bibnamefont {Kirtley}}, \bibinfo {author} {\bibfnamefont {C.}~\bibnamefont {Gould}}, \bibinfo {author} {\bibfnamefont {G.}~\bibnamefont {Tkachov}}, \bibinfo {author} {\bibfnamefont {E.~M.}\ \bibnamefont {Hankiewicz}}, \bibinfo {author} {\bibfnamefont {C.}~\bibnamefont {Br{\ifmmode\ddot{u}\else\"{u}\fi}ne}}, \bibinfo {author} {\bibfnamefont {H.}~\bibnamefont {Buhmann}}, \bibinfo {author} {\bibfnamefont {L.~W.}\ \bibnamefont {Molenkamp}}, \ and\ \bibinfo {author} {\bibfnamefont {K.~A.}\ \bibnamefont {Moler}},\ }\href {\doibase 10.1103/PhysRevLett.114.066801} {\bibfield  {journal} {\bibinfo  {journal} {Phys. Rev. Lett.}\ }\textbf {\bibinfo {volume} {114}},\ \bibinfo {pages} {066801} (\bibinfo {year} {2015})}\BibitemShut {NoStop}%
\bibitem [{\citenamefont {Ghatak}\ \emph {et~al.}(2018)\citenamefont {Ghatak}, \citenamefont {Breunig}, \citenamefont {Yang}, \citenamefont {Wang}, \citenamefont {Taskin},\ and\ \citenamefont {Ando}}]{Gathak2018Jul}%
  \BibitemOpen
  \bibfield  {author} {\bibinfo {author} {\bibfnamefont {S.}~\bibnamefont {Ghatak}}, \bibinfo {author} {\bibfnamefont {O.}~\bibnamefont {Breunig}}, \bibinfo {author} {\bibfnamefont {F.}~\bibnamefont {Yang}}, \bibinfo {author} {\bibfnamefont {Z.}~\bibnamefont {Wang}}, \bibinfo {author} {\bibfnamefont {A.~A.}\ \bibnamefont {Taskin}}, \ and\ \bibinfo {author} {\bibfnamefont {Y.}~\bibnamefont {Ando}},\ }\href {\doibase 10.1021/acs.nanolett.8b02029} {\bibfield  {journal} {\bibinfo  {journal} {Nano Lett.}\ }\textbf {\bibinfo {volume} {18}},\ \bibinfo {pages} {5124} (\bibinfo {year} {2018})}\BibitemShut {NoStop}%
\bibitem [{\citenamefont {Dam}\ \emph {et~al.}(2006)\citenamefont {Dam}, \citenamefont {Nazarov}, \citenamefont {Bakkers}, \citenamefont {Franceschi},\ and\ \citenamefont {Kouwenhoven}}]{vanDam2006Aug}%
  \BibitemOpen
  \bibfield  {author} {\bibinfo {author} {\bibfnamefont {J.~A.~v.}\ \bibnamefont {Dam}}, \bibinfo {author} {\bibfnamefont {Y.~V.}\ \bibnamefont {Nazarov}}, \bibinfo {author} {\bibfnamefont {E.~P. A.~M.}\ \bibnamefont {Bakkers}}, \bibinfo {author} {\bibfnamefont {S.~D.}\ \bibnamefont {Franceschi}}, \ and\ \bibinfo {author} {\bibfnamefont {L.~P.}\ \bibnamefont {Kouwenhoven}},\ }\href {\doibase 10.1038/nature05018} {\bibfield  {journal} {\bibinfo  {journal} {Nature}\ }\textbf {\bibinfo {volume} {442}},\ \bibinfo {pages} {667} (\bibinfo {year} {2006})}\BibitemShut {NoStop}%
\bibitem [{\citenamefont {Cleuziou}\ \emph {et~al.}(2006)\citenamefont {Cleuziou}, \citenamefont {Wernsdorfer}, \citenamefont {Bouchiat}, \citenamefont {Ondar{\ifmmode\mbox{\c{c}}\else\c{c}\fi}uhu},\ and\ \citenamefont {Monthioux}}]{Cleuziou2006Oct}%
  \BibitemOpen
  \bibfield  {author} {\bibinfo {author} {\bibfnamefont {J.-P.}\ \bibnamefont {Cleuziou}}, \bibinfo {author} {\bibfnamefont {W.}~\bibnamefont {Wernsdorfer}}, \bibinfo {author} {\bibfnamefont {V.}~\bibnamefont {Bouchiat}}, \bibinfo {author} {\bibfnamefont {T.}~\bibnamefont {Ondar{\ifmmode\mbox{\c{c}}\else\c{c}\fi}uhu}}, \ and\ \bibinfo {author} {\bibfnamefont {M.}~\bibnamefont {Monthioux}},\ }\href {\doibase 10.1038/nnano.2006.54} {\bibfield  {journal} {\bibinfo  {journal} {Nat. Nanotechnol.}\ }\textbf {\bibinfo {volume} {1}},\ \bibinfo {pages} {53} (\bibinfo {year} {2006})}\BibitemShut {NoStop}%
\bibitem [{\citenamefont {Kjaergaard}\ \emph {et~al.}(2017)\citenamefont {Kjaergaard}, \citenamefont {Suominen}, \citenamefont {Nowak}, \citenamefont {Akhmerov}, \citenamefont {Shabani}, \citenamefont {Palmstr{\o}m}, \citenamefont {Nichele},\ and\ \citenamefont {Marcus}}]{Kjaergaard2017Mar}%
  \BibitemOpen
  \bibfield  {author} {\bibinfo {author} {\bibfnamefont {M.}~\bibnamefont {Kjaergaard}}, \bibinfo {author} {\bibfnamefont {H.~J.}\ \bibnamefont {Suominen}}, \bibinfo {author} {\bibfnamefont {M.~P.}\ \bibnamefont {Nowak}}, \bibinfo {author} {\bibfnamefont {A.~R.}\ \bibnamefont {Akhmerov}}, \bibinfo {author} {\bibfnamefont {J.}~\bibnamefont {Shabani}}, \bibinfo {author} {\bibfnamefont {C.~J.}\ \bibnamefont {Palmstr{\o}m}}, \bibinfo {author} {\bibfnamefont {F.}~\bibnamefont {Nichele}}, \ and\ \bibinfo {author} {\bibfnamefont {C.~M.}\ \bibnamefont {Marcus}},\ }\href {\doibase 10.1103/PhysRevApplied.7.034029} {\bibfield  {journal} {\bibinfo  {journal} {Phys. Rev. Appl.}\ }\textbf {\bibinfo {volume} {7}},\ \bibinfo {pages} {034029} (\bibinfo {year} {2017})}\BibitemShut {NoStop}%
\bibitem [{\citenamefont {Goffman}\ \emph {et~al.}(2017)\citenamefont {Goffman}, \citenamefont {Urbina}, \citenamefont {Pothier}, \citenamefont {Nyg{\aa}rd}, \citenamefont {Marcus},\ and\ \citenamefont {Krogstrup}}]{Goffman2017Sep}%
  \BibitemOpen
  \bibfield  {author} {\bibinfo {author} {\bibfnamefont {M.~F.}\ \bibnamefont {Goffman}}, \bibinfo {author} {\bibfnamefont {C.}~\bibnamefont {Urbina}}, \bibinfo {author} {\bibfnamefont {H.}~\bibnamefont {Pothier}}, \bibinfo {author} {\bibfnamefont {J.}~\bibnamefont {Nyg{\aa}rd}}, \bibinfo {author} {\bibfnamefont {C.~M.}\ \bibnamefont {Marcus}}, \ and\ \bibinfo {author} {\bibfnamefont {P.}~\bibnamefont {Krogstrup}},\ }\href {\doibase 10.1088/1367-2630/aa7641} {\bibfield  {journal} {\bibinfo  {journal} {New J. Phys.}\ }\textbf {\bibinfo {volume} {19}},\ \bibinfo {pages} {092002} (\bibinfo {year} {2017})}\BibitemShut {NoStop}%
\bibitem [{\citenamefont {Hart}\ \emph {et~al.}(2019)\citenamefont {Hart}, \citenamefont {Cui}, \citenamefont {Ménard}, \citenamefont {Deng}, \citenamefont {Antipov}, \citenamefont {Lutchyn}, \citenamefont {Krogstrup}, \citenamefont {Marcus},\ and\ \citenamefont {Moler}}]{Hart2019Aug}%
  \BibitemOpen
  \bibfield  {author} {\bibinfo {author} {\bibfnamefont {S.}~\bibnamefont {Hart}}, \bibinfo {author} {\bibfnamefont {Z.}~\bibnamefont {Cui}}, \bibinfo {author} {\bibfnamefont {G.}~\bibnamefont {Ménard}}, \bibinfo {author} {\bibfnamefont {M.}~\bibnamefont {Deng}}, \bibinfo {author} {\bibfnamefont {A.~E.}\ \bibnamefont {Antipov}}, \bibinfo {author} {\bibfnamefont {R.~M.}\ \bibnamefont {Lutchyn}}, \bibinfo {author} {\bibfnamefont {P.}~\bibnamefont {Krogstrup}}, \bibinfo {author} {\bibfnamefont {C.~M.}\ \bibnamefont {Marcus}}, \ and\ \bibinfo {author} {\bibfnamefont {K.~A.}\ \bibnamefont {Moler}},\ }\href {\doibase 10.1103/physrevb.100.064523} {\bibfield  {journal} {\bibinfo  {journal} {Phys. Rev. B}\ }\textbf {\bibinfo {volume} {100}},\ \bibinfo {pages} {064523} (\bibinfo {year} {2019})}\BibitemShut {NoStop}%
\bibitem [{Sup()}]{Supplement}%
  \BibitemOpen
  \href@noop {} {}\bibinfo {note} {See Supplemental Material for sample preparation and measurement details, theoretical model description, and additional data.}\BibitemShut {Stop}%
\bibitem [{\citenamefont {Annunziata}\ \emph {et~al.}(2010)\citenamefont {Annunziata}, \citenamefont {Santavicca}, \citenamefont {Frunzio}, \citenamefont {Catelani}, \citenamefont {Rooks}, \citenamefont {Frydman},\ and\ \citenamefont {Prober}}]{Annunziata2010Oct}%
  \BibitemOpen
  \bibfield  {author} {\bibinfo {author} {\bibfnamefont {A.~J.}\ \bibnamefont {Annunziata}}, \bibinfo {author} {\bibfnamefont {D.~F.}\ \bibnamefont {Santavicca}}, \bibinfo {author} {\bibfnamefont {L.}~\bibnamefont {Frunzio}}, \bibinfo {author} {\bibfnamefont {G.}~\bibnamefont {Catelani}}, \bibinfo {author} {\bibfnamefont {M.~J.}\ \bibnamefont {Rooks}}, \bibinfo {author} {\bibfnamefont {A.}~\bibnamefont {Frydman}}, \ and\ \bibinfo {author} {\bibfnamefont {D.~E.}\ \bibnamefont {Prober}},\ }\href {\doibase 10.1088/0957-4484/21/44/445202} {\bibfield  {journal} {\bibinfo  {journal} {Nanotechnology}\ }\textbf {\bibinfo {volume} {21}},\ \bibinfo {pages} {445202} (\bibinfo {year} {2010})}\BibitemShut {NoStop}%
\bibitem [{\citenamefont {Kr{\o}jer}\ \emph {et~al.}(2024)\citenamefont {Kr{\o}jer}, \citenamefont {Dahl}, \citenamefont {Christensen}, \citenamefont {Kjaergaard},\ and\ \citenamefont {Flensberg}}]{Krojer2024Apr}%
  \BibitemOpen
  \bibfield  {author} {\bibinfo {author} {\bibfnamefont {S.}~\bibnamefont {Kr{\o}jer}}, \bibinfo {author} {\bibfnamefont {A.~E.}\ \bibnamefont {Dahl}}, \bibinfo {author} {\bibfnamefont {K.~S.}\ \bibnamefont {Christensen}}, \bibinfo {author} {\bibfnamefont {M.}~\bibnamefont {Kjaergaard}}, \ and\ \bibinfo {author} {\bibfnamefont {K.}~\bibnamefont {Flensberg}},\ }\href {\doibase 10.1103/PhysRevResearch.6.023064} {\bibfield  {journal} {\bibinfo  {journal} {Phys. Rev. Res.}\ }\textbf {\bibinfo {volume} {6}},\ \bibinfo {pages} {023064} (\bibinfo {year} {2024})}\BibitemShut {NoStop}%
\bibitem [{\citenamefont {Pillet}\ \emph {et~al.}(2019)\citenamefont {Pillet}, \citenamefont {Benzoni}, \citenamefont {Griesmar}, \citenamefont {Smirr},\ and\ \citenamefont {Girit}}]{Pillet2019Aug}%
  \BibitemOpen
  \bibfield  {author} {\bibinfo {author} {\bibfnamefont {J.-D.}\ \bibnamefont {Pillet}}, \bibinfo {author} {\bibfnamefont {V.}~\bibnamefont {Benzoni}}, \bibinfo {author} {\bibfnamefont {J.}~\bibnamefont {Griesmar}}, \bibinfo {author} {\bibfnamefont {J.-L.}\ \bibnamefont {Smirr}}, \ and\ \bibinfo {author} {\bibfnamefont {C.~O.}\ \bibnamefont {Girit}},\ }\href {\doibase 10.1021/acs.nanolett.9b02686} {\bibfield  {journal} {\bibinfo  {journal} {Nano Lett.}\ }\textbf {\bibinfo {volume} {19}},\ \bibinfo {pages} {7138} (\bibinfo {year} {2019})}\BibitemShut {NoStop}%
\bibitem [{\citenamefont {Melo}\ \emph {et~al.}(2022)\citenamefont {Melo}, \citenamefont {Fatemi},\ and\ \citenamefont {Akhmerov}}]{Melo2022Jan}%
  \BibitemOpen
  \bibfield  {author} {\bibinfo {author} {\bibfnamefont {A.}~\bibnamefont {Melo}}, \bibinfo {author} {\bibfnamefont {V.}~\bibnamefont {Fatemi}}, \ and\ \bibinfo {author} {\bibfnamefont {A.~R.}\ \bibnamefont {Akhmerov}},\ }\href {\doibase 10.21468/SciPostPhys.12.1.017} {\bibfield  {journal} {\bibinfo  {journal} {SciPost Phys.}\ }\textbf {\bibinfo {volume} {12}},\ \bibinfo {pages} {017} (\bibinfo {year} {2022})}\BibitemShut {NoStop}%
\bibitem [{\citenamefont {Matute-Cañadas}\ \emph {et~al.}(2023)\citenamefont {Matute-Cañadas}, \citenamefont {Tosi},\ and\ \citenamefont {Yeyati}}]{Matute2023Dec}%
  \BibitemOpen
  \bibfield  {author} {\bibinfo {author} {\bibfnamefont {F.~J.}\ \bibnamefont {Matute-Cañadas}}, \bibinfo {author} {\bibfnamefont {L.}~\bibnamefont {Tosi}}, \ and\ \bibinfo {author} {\bibfnamefont {A.~L.}\ \bibnamefont {Yeyati}},\ }\href {\doibase 10.48550/arXiv.2312.17305} {\bibfield  {journal} {\bibinfo  {journal} {arXiv}\ ,\ \bibinfo {pages} {2312.17305}} (\bibinfo {year} {2023})}\BibitemShut {NoStop}%
\bibitem [{\citenamefont {Zhang}\ \emph {et~al.}(2022)\citenamefont {Zhang}, \citenamefont {Gu}, \citenamefont {Li}, \citenamefont {Hu},\ and\ \citenamefont {Jiang}}]{Zhang2022Nov}%
  \BibitemOpen
  \bibfield  {author} {\bibinfo {author} {\bibfnamefont {Y.}~\bibnamefont {Zhang}}, \bibinfo {author} {\bibfnamefont {Y.}~\bibnamefont {Gu}}, \bibinfo {author} {\bibfnamefont {P.}~\bibnamefont {Li}}, \bibinfo {author} {\bibfnamefont {J.}~\bibnamefont {Hu}}, \ and\ \bibinfo {author} {\bibfnamefont {K.}~\bibnamefont {Jiang}},\ }\href {\doibase 10.1103/PhysRevX.12.041013} {\bibfield  {journal} {\bibinfo  {journal} {Phys. Rev. X}\ }\textbf {\bibinfo {volume} {12}},\ \bibinfo {pages} {041013} (\bibinfo {year} {2022})}\BibitemShut {NoStop}%
\end{thebibliography}%


\begin{thebibliography}{0}%
\makeatletter
\providecommand \@ifxundefined [1]{%
 \@ifx{#1\undefined}
}%
\providecommand \@ifnum [1]{%
 \ifnum #1\expandafter \@firstoftwo
 \else \expandafter \@secondoftwo
 \fi
}%
\providecommand \@ifx [1]{%
 \ifx #1\expandafter \@firstoftwo
 \else \expandafter \@secondoftwo
 \fi
}%
\providecommand \natexlab [1]{#1}%
\providecommand \enquote  [1]{``#1''}%
\providecommand \bibnamefont  [1]{#1}%
\providecommand \bibfnamefont [1]{#1}%
\providecommand \citenamefont [1]{#1}%
\providecommand \href@noop [0]{\@secondoftwo}%
\providecommand \href [0]{\begingroup \@sanitize@url \@href}%
\providecommand \@href[1]{\@@startlink{#1}\@@href}%
\providecommand \@@href[1]{\endgroup#1\@@endlink}%
\providecommand \@sanitize@url [0]{\catcode `\\12\catcode `\$12\catcode `\&12\catcode `\#12\catcode `\^12\catcode `\_12\catcode `\%12\relax}%
\providecommand \@@startlink[1]{}%
\providecommand \@@endlink[0]{}%
\providecommand \url  [0]{\begingroup\@sanitize@url \@url }%
\providecommand \@url [1]{\endgroup\@href {#1}{\urlprefix }}%
\providecommand \urlprefix  [0]{URL }%
\providecommand \Eprint [0]{\href }%
\providecommand \doibase [0]{http://dx.doi.org/}%
\providecommand \selectlanguage [0]{\@gobble}%
\providecommand \bibinfo  [0]{\@secondoftwo}%
\providecommand \bibfield  [0]{\@secondoftwo}%
\providecommand \translation [1]{[#1]}%
\providecommand \BibitemOpen [0]{}%
\providecommand \bibitemStop [0]{}%
\providecommand \bibitemNoStop [0]{.\EOS\space}%
\providecommand \EOS [0]{\spacefactor3000\relax}%
\providecommand \BibitemShut  [1]{\csname bibitem#1\endcsname}%
\let\auto@bib@innerbib\@empty
\end{thebibliography}%

\end{document}


\author{L.~Banszerus}
\affiliation{Center for Quantum Devices, Niels Bohr Institute, University of Copenhagen, 2100 Copenhagen, Denmark}%
\author{W. Marshall}
\affiliation{Center for Quantum Devices, Niels Bohr Institute, University of Copenhagen, 2100 Copenhagen, Denmark}
\affiliation{Department of Physics, University of Washington, Seattle, Washington 98195, USA}%
\author{C. W. Andersson}
\affiliation{Center for Quantum Devices, Niels Bohr Institute, University of Copenhagen, 2100 Copenhagen, Denmark}
\author{T. Lindemann}
\affiliation{Department of Physics and Astronomy, Purdue University, West Lafayette, Indiana 47907, USA}%
\affiliation{Birck Nanotechnology Center, Purdue University, West Lafayette, Indiana 47907, USA}
\author{M. J. Manfra}
\affiliation{Department of Physics and Astronomy, Purdue University, West Lafayette, Indiana 47907, USA}%
\affiliation{Birck Nanotechnology Center, Purdue University, West Lafayette, Indiana 47907, USA}
\affiliation{School of Electrical and Computer Engineering, Purdue University, West Lafayette, Indiana 47907, USA}
\affiliation{School of Materials Engineering, Purdue University, West Lafayette, Indiana 47907, USA}
\author{C. M. Marcus}
\affiliation{Center for Quantum Devices, Niels Bohr Institute, University of Copenhagen, 2100 Copenhagen, Denmark}
\affiliation{Department of Physics, University of Washington, Seattle, Washington 98195, USA}%
\affiliation{Materials Science and Engineering, University of Washington, Seattle, Washington 98195, USA}%
\author{S. Vaitiek\.{e}nas}
\affiliation{Center for Quantum Devices, Niels Bohr Institute, University of Copenhagen, 2100 Copenhagen, Denmark}%

\title{Supplemental Material:\\ Voltage-controlled synthesis of higher harmonics in hybrid Josephson junction circuits}

\date{\today}

\maketitle

\section*{Sample preparation}
The devices were fabricated on a semiconductor-superconductor hybrid heterostructure. The III-V semiconductor stack was grown on a semi-insulating InP substrate by molecular beam epitaxy (MBE). It consists of a 7~nm thick InAs quantum well, encapsulated between a 4~nm thick In$_{0.75}$Ga$_{0.25}$As bottom barrier from below and a 10~nm thick In$_{0.75}$Ga$_{0.25}$As top barrier, followed by two monolayers of GaAs capping.
The 5~nm Al film was grown \textit{in situ} without exposing the heterostructure to ambient conditions. 

The devices were fabricated using standard electron beam lithography (Elionix, 100~keV). The Al was selectively etched using Transene Aluminum etch type~D at 50~$^\circ$C for 5 seconds. The III-V heterostructure was structured by a  chemical wet etch using (220:55:3:3 H$_2$O:C$_6$H$_8$O$_7$:H$_3$PO$_4$:H$_2$O$_2$). The HfOx (18~nm) gate dielectrics were grown using atomic layer deposition (Veeco Savannah). The first Ti/Au (3/20 nm) and the second Ti/Au (5/350~nm) gate layers were deposited using electron beam evaporation (AJA International) at a base pressure of $10^{-8}$~mbar.
The gate above $J_3$ showed no response, presumably due to unsuccessful fabrication, and was kept at $V_3 = 0$. The global top gate was kept at $-1.3$~V throughout the experiment.

The electrical properties of the two-dimensional electron gas were characterized in a gated Hall bar geometry, where the Al film has been removed. The measured charge carrier mobility peaked at $\mu=53000$~cm$^{2}$/Vs for a charge carrier density of $n=0.6\times 10^{12}~$cm$^{-2}$/Vs.
Normal state sheet resistance, $R_{\square} = 6.5~\Omega$, used for kinetic inductance estimations, was measured separately above the Al critical temperature in a Hall bar geometry where the Al film was not removed.
Tunneling spectroscopy measurements showed an induced superconducting gap of $\Delta=180~\mu$eV.

\section*{Measurements}
Electrical transport measurements were performed using a commercial cryofree dilution refrigerator (Oxford Instruments, Triton) at a base temperature of 20~mK.
Standard dc and low-frequency ac ($f=27.4$~Hz) lock-in measurement techniques were used.
All lines were filtered at cryogenic temperatures using a commercial two-stage RC and LC filter with a cut-off frequency of 65~kHz (QDevil, QFilter).
Gate lines were additionally filtered using 16~Hz low pass filters at room temperature.
The dc and ac currents were amplified using a commercially available current-to-voltage converter (Basel Precision Instruments) using a gain of $10^{6}$.
Both dc and ac voltages were amplified using commercial preamplifiers (Stanford Research, SR560) at a gain of $10^{3}$.

\section*{Calculation of higher harmonics}

The ratio of the first and second harmonic, $A_\mathrm{4e}/A_\mathrm{2e}$, can be analytically calculated by expressing  Eq.~(2) of the main text as a Fourier series. Since $I(\varphi)$ is periodic on the interval $-\pi < \varphi < \pi$ and an odd function of $\varphi$, its Fourier series representation is given by

\begin{equation}
    I(\varphi) \sim \frac{e\sigma}{2\hbar}\sum_{n=1}^{\infty}A_{(2n)e}\sin(n\varphi).
\end{equation}

\noindent The $n^{\rm th}$ harmonic coefficient, $A_{(2n)e}$, associated with $(2n)e$ charge transport, is given by

\begin{equation}
    A_{(2n)e} \equiv \frac{2}{\pi}\int_0^{\pi} \frac{\tau \sin\varphi}{\sqrt{1-\tau\sin^2\left (\varphi/2\right)}}\sin(n\varphi) \ d\varphi,
\end{equation}

\noindent where \mbox{$\tau=4\rho/(1+\rho)^2$} is the effective transparency of the double junction. The coefficient $A_{(2n)e}$ can be expressed in terms of the complete elliptic integrals of the first and second kind, $K(x)$ and $E(x)$, respectively, given by

\begin{align}
    K(\tau)&=\int_0^{\pi/2}\frac{d\varphi}{\sqrt{1-\tau \sin^2 \varphi}},
    \\  \nonumber
    \\
    E(\tau) &= \int_0^{\pi/2} \sqrt{1-\tau \sin^2 \varphi} \ d\varphi.
\end{align}

\noindent The ratio of the second coefficient to the first is given by
\begin{equation*}
    \frac{A_\mathrm{4e}}{A_\mathrm{2e}}=  \frac{2}{5} \frac{\left [8\left ( -1+\frac{3}{\tau} - \frac{2}{\tau^2}\right ) K(\tau) + \left ( 1-\frac{16}{\tau} +\frac{16}{\tau^2} \right ) E(\tau)  \right ]}{\left [2\left ( 1-\frac{1}{\tau} \right ) K(\tau) + \left ( \frac{2}{\tau} -1 \right ) E(\tau)  \right ] }\,. 
\end{equation*}

\noindent The ratio of coefficients, $A_\mathrm{4e}/A_\mathrm{2e}$, shown in Fig.~5 of the main text and Fig.~\ref{fs4} was then determined using values for $\rho$ obtained by fitting Eq.~(2) in the main text to the measured CPR, as shown in Fig.~4 of the main text.\\ \\

\section*{Model with finite junction transparency}

The analytical model described in the main text considers two sinusoidal JJs in series.
In practice, however, hybrid JJs based on high-mobility quantum wells might exhibit a sizable intrinsic transparency, $\tau_\mathrm{int}$, leading to a nonsinusoidal CPR in the individual junctions, given by 
\begin{equation}
    I(\varphi)=\frac{e\Delta}{2\hbar}\frac{\tau_\mathrm{int}\,\sin(\varphi)}{\sqrt{1-\tau_\mathrm{int}\,\sin^2(\varphi/2)}}\,.
    \label{Beenakker}
\end{equation}
To study the effect of a finite $\tau_\mathrm{int}$ of the individual JJs on the CPR of the two JJs in series, we model the Josephson energy of each junction using the single-mode expression with finite transparency, $E_i(\varphi_i)=-E_{\mathrm{J}i}\sqrt{1-\tau_\mathrm{int}\sin^2(\varphi_i/2)}$, and assume $\tau_\mathrm{int}$ to be identical in the two JJs.
We then follow the same steps as for the two sinusoidal JJs discussed in the main text.
We derive the ground-state energy, $E_0$, by numerically minimizing the total energy, $E_1(\varphi_1) + E_2(\varphi_2)$, under the boundary condition of the total phase bias, $\varphi=\varphi_1+\varphi_2$.
The corresponding CPR is calculated by numerically differentiating the ground state energy, $ I(\varphi) =\frac{2e}{\hbar}\frac{\partial E_\mathrm{0}}{\partial \varphi}$.
We find that the modeled CPR of a symmetrized double JJ ($\rho=1$) depends only weakly on $\tau_\mathrm{int}$; see Fig.~\ref{fs3}(a). 
To better understand this behavior, we investigate fast Fourier spectra of the modeled CPRs for different combinations of $\rho$ and $\tau_\mathrm{int}$.
We find that two sinusoidal and symmetrized JJs in series [purple bars in Fig.~\ref{fs3}(b)] have nearly the same harmonic content as a strongly desymmetrized double JJ with $\tau_\mathrm{int}=1$ [orange bars in Fig.~\ref{fs3}(b)].
We note that the harmonic composition of two symmetrized JJs, each with unity transparency, displays only slightly larger amplitudes of the higher harmonics compared to the other two cases [yellow bars in Fig~\ref{fs3}(b)].

To further explore the effect of finite intrinsic transparencies on the harmonic composition, we calculate the ratio of second to first harmonic amplitudes, $A_\mathrm{4e}/A_\mathrm{2e}$, as a function of $\rho$ and $\tau_\mathrm{int}$ [Fig.~\ref{fs3}(c)].
We observe that a large $A_\mathrm{4e}/A_\mathrm{2e}$ can be obtained either by $\rho \approx 1$ or by $\tau_\mathrm{int} \approx 1$. 
To obtain a sinusoidal CPR, both $\rho$ and $\tau_\mathrm{int}$ have to be close to zero; however, a finite $\tau_\mathrm{int}$ limits how sinusoidal the CPR can become when desymmetrizing the two JJs [see Fig.~\ref{fs3}(d)].
We find that the model of two sinusoidal JJs holds well for $\tau_\mathrm{int} < \rho/(1+\rho)^2$.
Experimentally, we observe that $\tau_\mathrm{int}\rightarrow 0$ is a good approximation, except for small $\rho$ in arm 2 [see Fig.~\ref{fs4}], where the measured $A_\mathrm{4e}/A_\mathrm{2e}$ exceeds the modeled value for two sinusoidal JJs (dashed red line) and aligns better with a model of two JJs with $\tau_\mathrm{int}=0.5$ (dashed green line).

\onecolumngrid
\vspace{2cm}
\newpage

\begin{figure}[h]
    \includegraphics[width=0.5\linewidth]{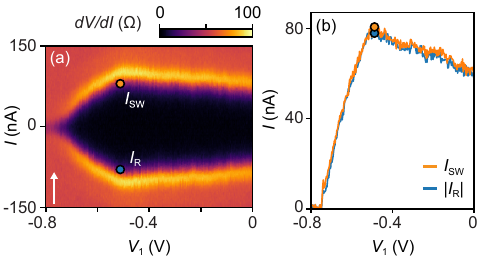}
    \caption{
    (a) Differential resistance, $dV/dI$, as a function of current bias, $I$, and $J_1$ gate voltage $V_\mathrm{1}$, measured for arm 1 with  open $J_\mathrm{2}$ ($V_2 = 0$).
    Arm 2 and $J_\mathrm{ref}$ were pinched off ($V_\mathrm{4} = -1$~V, $V_{\rm ref} = - 1.5$~V).
    The measurements were taken by sweeping $I$ from negative to positive, as indicated by the white arrow.
    (b) The magnitude of the switching, $I_\mathrm{SW}$, and retrapping, $I_\mathrm{R}$, currents as a function of $V_\mathrm{1}$ show a nonhysteretic behavior, characteristic for an overdamped junction.
    $I_{\rm SW}$ and $I_{\rm R}$ were extracted via the threshold method, using a threshold of $30~\Omega$, about half of the normal-state resistance.
    }
    \label{fs1}
\end{figure}


\begin{figure}[h!]
    \includegraphics[width=0.5\linewidth]{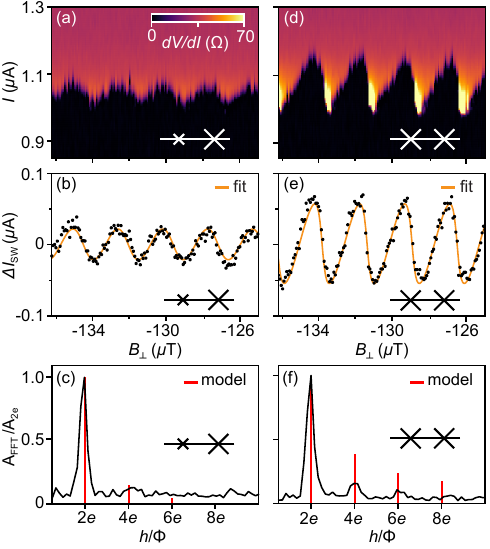}
    \caption{
    (a) Differential resistance, $dV/dI$, as a function of current bias, $I$, and perpendicular magnetic field, $B_\perp$, for arm~1 in the desymmetrized regime ($V_1=-0.6~V$, $V_2=-0.7~V$) with open $J_{\rm ref}$ ($V_\mathrm{ref}=0$) but pinched-off arm~2 ($V_3=0~V$, $V_4=-1~V$).
    (b) Switching current oscillations, $\Delta I_\mathrm{SW}$, inferred from data in (a). Fit to the main-text Eq.~(2) yields $\rho=0.14$ corresponding to an effective $\tau=0.42$. 
    (c) Fast Fourier transform amplitude of $\Delta I_\mathrm{SW}$ from (b), in the range of $B_\perp$=-140 to -110 $\mu T$. The red bars indicate the expected amplitude of the harmonics calculated from the ratio of the switching currents of $J_1$ and $J_2$.    
    (d)--(f)~Same as (a)--(c) but for the symmetrized case with 
    $V_1=-0.4$~V, $V_2=-0.45$~V.
    The fit to the main-text Eq.~(2) in (e) yields $\rho=0.54$, corresponding to an effective $\tau=0.91$.
    }
    \label{fs2}
\end{figure}

\begin{figure}[t]
    \includegraphics[width=0.7\linewidth]{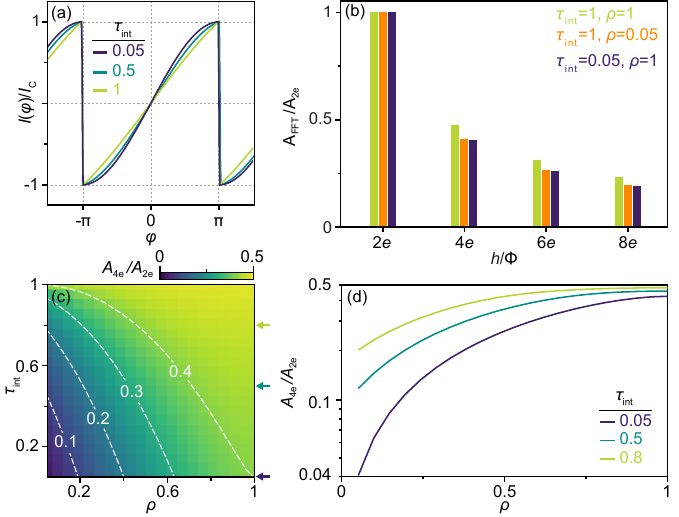}
    \caption{
    (a)~Modeled current-phase relation of two JJs in series with equal Josephson energies ($\rho=1$). The intrinsic transparency of the individual junctions, $\tau_\mathrm{int}$, varies between low (sinusoidal) and high (nonsinusoidal) limits.
    (b)~Fast Fourier spectra of the model CPRs for two sinusoidal and symmetrized  (purple), nonsinusoidal and desymmetrized (orange), and nonsinusoidal and symmetrized (yellow) JJs in series. 
    (c)~Calculated ratio of second to first harmonic amplitudes, $A_\mathrm{4e}/A_\mathrm{2e}$, as a function of $\tau_\mathrm{int}$ and $\rho$.
    (d)~Line cuts of $A_\mathrm{4e}/A_\mathrm{2e}$ taken from (c) as a function of $\rho$ at $\tau_\mathrm{int} = 0.05,\ 0.5,$ and $0.8$.
     }
    \label{fs3}
\end{figure}

\begin{figure}[t]
    \includegraphics[width=0.5\linewidth]{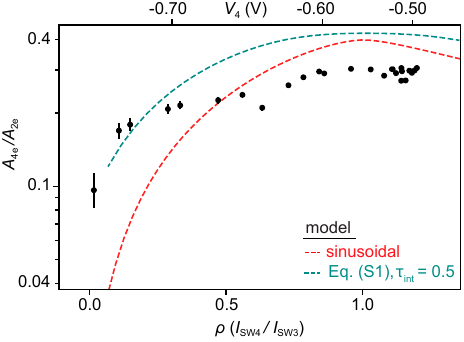}
    \caption{
    Ratio of second to first harmonic amplitudes, $A_\mathrm{4e}/A_\mathrm{2e}$, determined from fits to CPR, plotted against independently measured $\rho=I_\mathrm{SW4}/I_\mathrm{SW3}$ for arm 2.
    Error bars indicate uncertainties in the fits to the main-text Eq.~(2).
    Data were taken with varying $V_4$ between $-0.75$ and $-0.5$~V and fixed $V_3=0$~V.
    The dashed red curve is the expected $A_\mathrm{4e}/A_\mathrm{2e}$ based on the measured switching currents, assuming two sinusoidal JJs. The green dashed curve represents $A_\mathrm{4e}/A_\mathrm{2e}$ based on a model that considers a finite intrinsic JJ transparency, $\tau_\mathrm{int}=0.5$, for each JJ.
    }
    \label{fs4}
\end{figure}